\documentclass[aps,pra,reprint,showpacs,notitlepage,superscriptaddress,twocolumn]{revtex4-2}
\usepackage{amssymb}
\usepackage{mathrsfs}
\usepackage{amsfonts}
\usepackage{graphicx}
\usepackage{amsmath}
\usepackage{dcolumn}
\usepackage{color}
\usepackage{subfigure}
\usepackage{epsfig}
\usepackage{soul}
\usepackage[usenames,dvipsnames]{xcolor}
\usepackage[colorlinks,linkcolor=blue,citecolor=blue,hyperindex,bookmarks=false,pdfstartview=FitH]{hyperref}
\usepackage{bm}
\usepackage{dsfont}
\usepackage{verbatim}
\usepackage{multirow}
\usepackage{mathrsfs}
\usepackage{leftidx}
\usepackage{xspace}
\usepackage{braket}
\usepackage{bbm}
\usepackage{cancel}
\usepackage[normalem]{ulem}
\providecommand{\U}[1]{\protect\rule{.1in}{.1in}}

\newcommand{\figpanel}[2]{\hyperref[#1]{\ref*{#1}(#2)}}
\begin{document}

\title{Spinning giant optomechanical cavity with nonreciprocal self-interference}
\author{Yao-Tong Chen}
\address{ICFO – Institut de Ciencies Fotoniques, The Barcelona Institute of Science and Technology, Castelldefels, Barcelona 08860, Spain}

\date{\today}

\begin{abstract}
We study a spinning optomechanical cavity that is coupled to a meandering waveguide at multiple spatially separated points, forming a giant-cavity configuration. 
The resulting self-interference makes the effective optical driving, linewidth, and frequency shift strongly dependent on the propagation phase in the waveguide, while the rotation-induced Sagnac--Fizeau shift causes the clockwise (CW) and counterclockwise (CCW) cavity modes to experience distinct interference phases.
Under single-tone driving, this mechanism enables phase-controlled phonon cooling and, in the presence of cavity rotation, nonreciprocal cooling, with one propagation direction approaching the mechanical ground-state regime while the opposite direction remains less efficiently cooled.
Under two-tone driving, the same interference mechanism engineers a squeezed reservoir for the mechanical mode, producing steady-state squeezing that likewise becomes nonreciprocal under cavity rotation. 
These results establish multi-point self-interference as a versatile mechanism for phase-controlled optomechanical reservoir engineering and show that, when combined with cavity rotation, it provides a route to nonreciprocal quantum effects.
\end{abstract}

\maketitle

\section{Introduction}
\label{sec:introduction}

Optical nonreciprocity, in which the response of a system depends on the propagation direction of light, is a key resource for signal routing, isolation, and directional control of classical and quantum fields~\cite{Jalas2013,Lodahl2017,Caloz2018}. 
In optomechanical systems, optical nonreciprocity has been realized by engineering the interactions and interference pathways between optical and mechanical modes~\cite{Manipatruni2009PRL,Ruesink2016,Fang2017NatPhys}.
A conceptually different route is to physically rotate a whispering-gallery-mode resonator, for which the Sagnac--Fizeau effect shifts the clockwise (CW) and counterclockwise (CCW) resonances in opposite directions~\cite{Lu2017PRJ,Maayani2018Nature}. 
Consequently, fields incident from opposite directions experience different effective detunings and therefore different optical responses. 
This mechanism has enabled a variety of classical nonreciprocal effects, including optical isolation~\cite{Maayani2018Nature}, direction-dependent optomechanically induced transparency~\cite{Lu2017PRJ}, unidirectional phonon lasing~\cite{Jiang2018PRAppl}, and nonreciprocal chaotic dynamics~\cite{Zhang2021PRA}. 
It has further enabled various nonreciprocal quantum effects, including photon~\cite{Huang2018PRL} and phonon~\cite{Yuan2023OE} blockade, optomechanical entanglement~\cite{Jiao2020PRL}, and mechanical squeezing~\cite{Guo2023PRA}. 
In most of these spinning-cavity schemes, the Sagnac--Fizeau shift enters the optical dynamics primarily as a direction-dependent detuning, such that the nonreciprocity originates from different resonance conditions experienced by the two propagation directions.

\begin{figure}[t]
\centering
\includegraphics[width=0.44\textwidth]{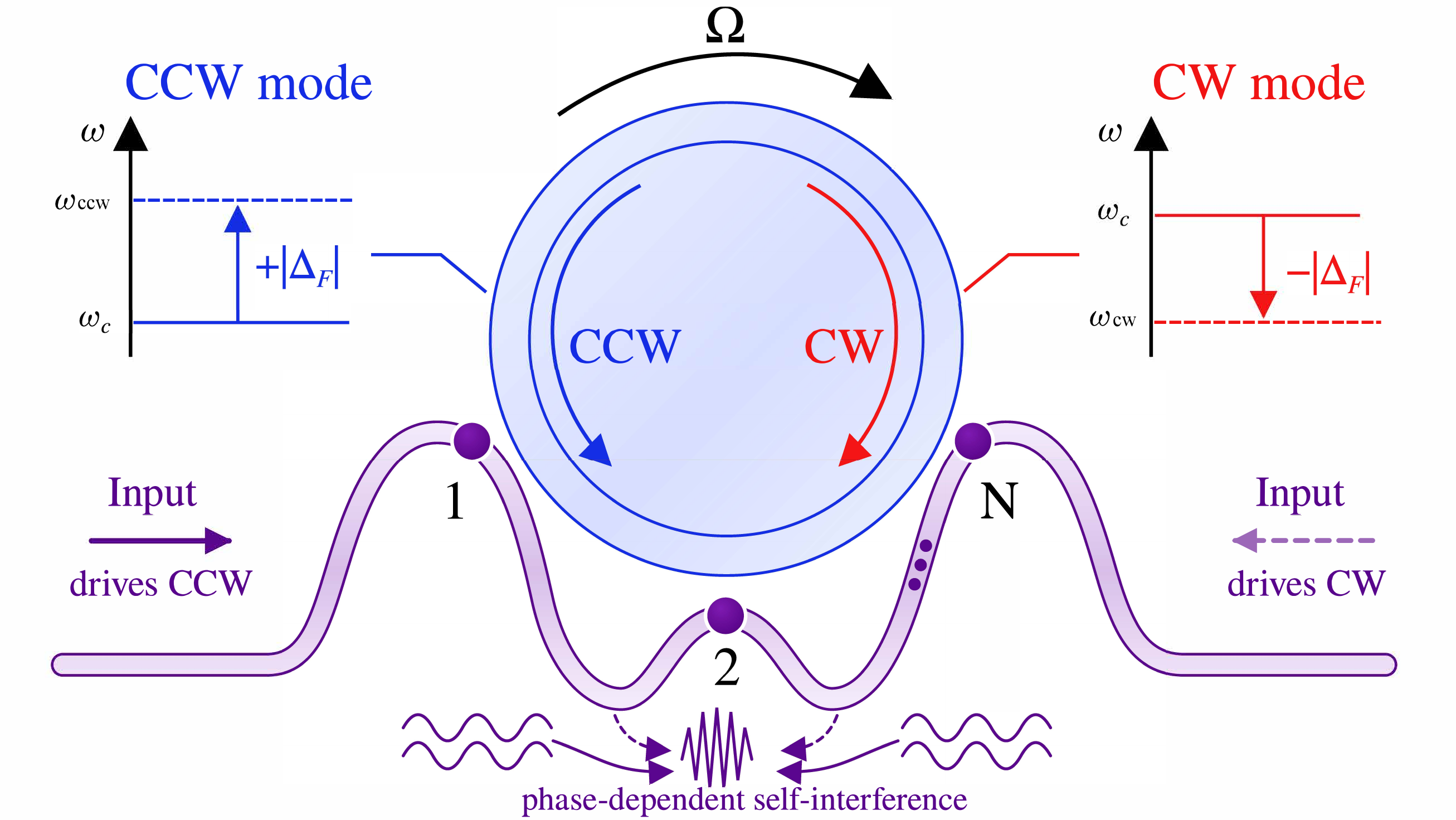}
\caption{
Schematic of a giant spinning optomechanical cavity. 
A spinning ring cavity is evanescently coupled to a meandering waveguide at multiple spatially separated points, forming a giant-cavity configuration.
The cavity supports two counterpropagating optical modes, denoted by CW and CCW, whose resonance frequencies are shifted in opposite directions by the rotation at angular velocity $\Omega$.
The multi-point waveguide coupling leads to self-interference, which makes the effective detuning, dissipation, and driving phase dependent and direction dependent.
}
\label{model}
\end{figure}

In this paper, we pursue a different route by combining the rotation-induced frequency splitting of a spinning optomechanical cavity with multi-point self-interference through a waveguide. 
Our construction is inspired by the giant-atom paradigm in waveguide quantum electrodynamics (QED)~\cite{anton2014,Gustafsson2014,FriskKockum2020}, where an artificial atom couples to a photonic or phononic environment at multiple spatially separated points~\cite{Kannan2020}.
The propagation phases accumulated between these coupling points give rise to interference among the different coupling channels, leading to strongly phase-dependent relaxation rates and Lamb shifts~\cite{anton2014}. 
This interference mechanism has also been exploited in giant-atom waveguide QED to control single-photon scattering~\cite{Wang2019,Du2021,Cai2021,Du2021prr,Chen2022} and engineer exotic bound states~\cite{Guo2020,Zhao2020,Wang2021,Vega2021,Xiao2022,Lim2023}.
When the propagation delays between different coupling points become non-negligible,
giant atoms can further exhibit non-Markovian 
dynamics~\cite{Guo2017,Andersson2019,Du2022}.
Such multi-point coupling structures have been demonstrated experimentally not only with superconducting artificial atoms~\cite{Kannan2020}, but also with ferromagnetic spin ensembles~\cite{Wang2022,Wang2026}. 
The underlying concept has subsequently been extended from atomic emitters to bosonic resonators, where multi-point self-interference enables tunable optical responses and enhanced sensing performance~\cite{Wan2018,Du2021OE,Zhu2022FrontPhys}. 
For instance, a spinning giant cavity, which is spatially extended through its coupling to a meandering waveguide at multiple points, has been shown to support chiral emission and nonreciprocal single-photon transport~\cite{Liu2022}.
More recently, a nonlinear giant cavity has been explored for controlling chiral multiphoton dynamics, direction-dependent photon statistics, and dissipative phase transitions~\cite{Chang2025}. 

Motivated by these developments, we consider a spinning
whispering-gallery-mode optomechanical cavity coupled to a meandering
waveguide at multiple spatially separated points, thereby forming a giant-cavity system. The resulting self-interference
simultaneously modifies the effective coherent drive, optical
linewidth, and interference-induced frequency shift. The
Sagnac--Fizeau splitting shifts the CW and CCW resonances, so that
the two counterpropagating modes acquire different interference phases and therefore experience
nonreciprocal engineered optical reservoirs. Importantly, the nonreciprocity in our scheme does not simply arise from a frequency-axis displacement of the CW and CCW response spectra. 
Rather, it can be understood from direction-dependent self-interference, which leads to genuinely different effective optical reservoirs for the two propagation directions.
We show that this mechanism enables phase-controlled sideband cooling under single-tone driving and reservoir-engineered mechanical squeezing under two-tone driving, both of which become nonreciprocal when combined with cavity
rotation.

This paper is organized as follows. 
In Sec.~\ref{sec:general_model}, we introduce the giant spinning optomechanical cavity and derive the phase-dependent effective driving amplitude, optical dissipation, and frequency shift generated by the multi-point waveguide coupling. 
In Sec.~\ref{sec:cooling_theory}, we develop the linearized covariance-matrix description of single-tone sideband cooling, and in Sec.~\ref{sec:cooling_results} we study phase-dependent and nonreciprocal phonon cooling. 
In Sec.~\ref{sec:squeezing_theory}, we formulate the two-tone reservoir-engineering scheme and derive the mechanical squeezing within the rotating-wave approximation, while Sec.~\ref{sec:squeezing_results} presents the corresponding phase-dependent and nonreciprocal squeezing results.

\section{Giant cavity model}
\label{sec:general_model}

We consider a spinning whispering-gallery-mode optomechanical cavity evanescently coupled to a meandering optical fiber at multiple spatially separated points, as shown in Fig.~\ref{model}. For simplicity, the coupling points are assumed to be equally spaced along the propagation path, so that the same phase is accumulated between two neighboring points. Due to the rotation of the cavity, the two counterpropagating optical modes, namely the clockwise (CW) and counterclockwise (CCW) modes, acquire different resonance frequencies via the Sagnac--Fizeau shift. The corresponding direction-dependent frequency shift is
\begin{equation}
\Delta_{F}
=
\pm \Omega
\frac{n_1R\omega_{c}}{c}
\left(
1-\frac{1}{n_1^{2}}
-\frac{\lambda}{n_1}\frac{dn_1}{d\lambda}
\right),
\label{eq:Fizeau}
\end{equation}
where $\Omega$ is the angular velocity of the cavity, $R$ is the cavity radius, $n_1$ is the refractive index, $\omega_c$ is the resonance frequency of the static cavity, and $\lambda$ is the vacuum wavelength. 
For the parameter regime considered here, the material-dispersion term $dn_1/d\lambda$ is much smaller than the leading contribution and will be neglected.
For a given CW rotation with $\Omega>0$, the effective resonance frequencies are
\begin{equation}
\omega_{\rm cw}=\omega_c-|\Delta_F|,
\quad
\omega_{\rm ccw}=\omega_c+|\Delta_F|.
\label{eq:cw_ccw_freq}
\end{equation}
This rotation-induced frequency splitting breaks the degeneracy between the two counterpropagating modes and, when combined with the phase-dependent multi-point interference introduced below, provides the basis for nonreciprocal optomechanical responses.

In addition, the resonator supports a mechanical breathing mode with frequency $\omega_m$. The system Hamiltonian reads
\begin{equation}
\begin{split}
H
=
&\hbar\sum_{j={\rm cw,ccw}}
\omega_j \hat a_j^\dagger \hat a_j
+
\frac{\hbar\omega_m}{2}
\left(
\hat q^2+\hat p^2
\right)
\\
&-
\hbar g
\sum_{j={\rm cw,ccw}}
\hat a_j^\dagger \hat a_j \hat q
+
H_d ,
\label{eq:H_general}
\end{split}
\end{equation}
where $\hat a_j$ annihilates a photon in the optical mode $j$, $\hat q$ and $\hat p$ are dimensionless mechanical quadratures satisfying $[\hat q,\hat p]=i$, and $g$ is the single-photon optomechanical coupling rate.
The form of the driving Hamiltonian $H_d$ depends on the driving scheme considered below.
In the cooling configuration, a single  red-detuned tone is applied. 
In the squeezing configuration, two coherent tones are applied near the red and blue mechanical sidebands. 
The same multi-point interference structure underlies both configurations.

The input direction selects which counterpropagating optical mode is driven. 
When the field is incident from the left, the CCW mode is excited; when the field is incident from the right, the CW mode is excited.
Accordingly, we take $j={\rm ccw}$ for left driving and $j={\rm cw}$ for right driving. For a fixed input direction, the opposite traveling-wave mode is not directly driven.
In the absence of appreciable Rayleigh-scattering-induced backscattering between the CW and CCW modes~\cite{Mazzei2007}, its mean photon number vanishes in the linearized treatment.
Therefore, only the driven mode is retained explicitly in the reduced linearized fluctuation equations.

We now formulate the self-interference induced by the multi-point waveguide coupling.
For the mode $j$, we denote by $\phi_j$ the propagation phase accumulated between two neighboring coupling points.
Within the standard input-output formalism, the field at the $s$th coupling point obeys~\cite{Gardiner1985}
\begin{subequations}
\label{eq:io_multiport}
\begin{align}
\hat a_{s,j}^{\rm out}
&=\hat a_{s,j}^{\rm in}- 
\sqrt{2\kappa_{{\rm ex},j}}\hat a_j
,
\\
\hat a_{s+1,j}^{\rm in}
&=
e^{i\phi_j}\hat a_{s,j}^{\rm out}.
\end{align}
\end{subequations}
Here $s=1,\ldots,N$ labels the coupling points and $\kappa_{{\rm ex},j}$ denotes the local external coupling rate, taken to be the same for all points.
The ordering of the index $s$ follows the propagation direction of the incident field.

The contribution from the $N$ coupling points is captured by the interference amplitude~\cite{anton2014}
\begin{equation}
\mathcal A_j(\phi_j)
=\sum_{s=1}^{N}
e^{i(N-s)\phi_j}.
\label{eq:A_phi_general}
\end{equation}
The effective driving amplitude of a coherent input field with amplitude $E$ is then
\begin{equation}
E_{{\rm eff},j}
=
E\mathcal A_j(\phi_j).
\label{eq:Eeff_general}
\end{equation}
The same multi-point interference produces a complex optical self-energy, whose real and imaginary parts describe the effective dissipation and frequency shift, respectively. For identical local coupling rates, this gives
\begin{equation}
\kappa_{{\rm eff},j}(\phi_j)
=
N\kappa_{{\rm ex},j}
+
2\kappa_{{\rm ex},j}
\sum_{s=1}^{N-1}
(N-s)\cos(s\phi_j),
\label{eq:kappaeff_general}
\end{equation}
and
\begin{equation}
\Delta_{{\rm eff},j}(\phi_j)
=
2\kappa_{{\rm ex},j}
\sum_{s=1}^{N-1}
(N-s)\sin(s\phi_j).
\label{eq:Deltaeff_general}
\end{equation}
The noise entering through the multiple coupling points can also be written as a coherent superposition,
\begin{equation}
\sum_{s=1}^{N}
\sqrt{2\kappa_{{\rm ex},j}}\hat a_{s,j}^{\rm in}
=
\sqrt{2\kappa_{{\rm ex},j}}
\sum_{s=1}^{N}
e^{i(N-s)\phi_j}
\hat a_j^{\rm in}.
\label{eq:noise_general}
\end{equation}
Here $\hat a_{s,j}^{\rm in}$ denotes the input noise arriving at the $s$th coupling point for mode $j$.
Since all coupling points are connected by the same traveling waveguide field, the noise operators $\hat a_{s,j}^{\rm in}$ at different coupling points are not independent reservoirs, but are related by the propagation phases accumulated along the waveguide. 
We choose $\hat a_j^{\rm in}$ as a reference input noise operator of this traveling vacuum field with zero mean and Markovian correlation $\left\langle
\hat a_j^{\rm in}(t)
\hat a_{j'}^{{\rm in},\dagger}(t')
\right\rangle
=
\delta_{j j'}\delta(t-t')$. 
As a result, the optical diffusion associated with the external coupling channel is
\begin{equation}
F_j(\phi_j)
=
\kappa_{{\rm ex},j}
\left|
\mathcal A_j(\phi_j)
\right|^2 .
\label{eq:F_general}
\end{equation}
Thus, for the external waveguide channel, $F_j(\phi_j)=\kappa_{{\rm eff},j}(\phi_j)$. 
These phase-dependent quantities provide the common building blocks for the cooling and squeezing calculations below.

\section{Sideband cooling}
\label{sec:cooling_theory}

We now specialize the general model to the sideband-cooling
configuration~\cite{WilsonRae2007,Marquardt2007,Aspelmeyer2014}.
Resolved-sideband cooling and near-ground-state preparation have
been demonstrated in both optical and microwave optomechanical
platforms~\cite{Schliesser2008,Teufel2011,Chan2011}.
A single red-detuned tone with frequency $\omega_{d,{\rm c}}$ is applied to the selected optical mode $j$. The driving Hamiltonian is
\begin{equation}
H_{d,{\rm c}}
=
i\hbar
\left[
E_{{\rm eff},j}\hat a_j^\dagger e^{-i\omega_{d,{\rm c}} t}
-
E_{{\rm eff},j}^{*}\hat a_j e^{i\omega_{d,{\rm c}} t}
\right],
\label{eq:Hd_cooling}
\end{equation}
where the subscript ``c'' denotes the cooling configuration. 
The incident drive amplitude is related to the input power by
$E=\sqrt{2\kappa_{{\rm ex},j}P/(\hbar\omega_c)}$.
In the rotating frame of the drive, the quantum Langevin equations are
\begin{subequations}
\label{QLE}
\begin{align}
\dot{\hat{a}}_{j}
=&-(i\Delta_{j}+\kappa_{j})\hat{a}_{j}
+ig\hat{a}_{j}\hat{q}
+E_{\text{eff},j} \notag\\
&+\sqrt{2\kappa_{0}}\hat{a}_{0,j}^{\text{in}}
+\sum_{s=1}^{N}\sqrt{2\kappa_{\text{ex},j}}\hat{a}_{s,j}^{\text{in}},
\label{QLEa}
\\
\dot{\hat{q}}
=&\omega_{m}\hat{p},
\label{QLEb}
\\
\dot{\hat{p}}
=&-\omega_{m}\hat{q}
-\gamma_{m}\hat{p}
+g\hat{a}_{j}^{\dagger}\hat{a}_{j}
+\hat{\xi}.
\label{QLEc}
\end{align}
\end{subequations}
Here $\gamma_m$ is the mechanical damping rate. 
In the Markovian approximation for the mechanical bath, the Brownian noise $\hat \xi$ satisfies~\cite{Aspelmeyer2014} $\frac{1}{2}
\left\langle
\hat\xi(t)\hat\xi(t')
+
\hat\xi(t')\hat\xi(t)
\right\rangle
\simeq
\gamma_m(2n_m+1)\delta(t-t')$, 
where $n_m=[\exp(\hbar\omega_m/k_BT)-1]^{-1}$ is the thermal phonon occupation, $k_B$ is the Boltzmann constant, and $T$ is the environment temperature.
The effective cavity decay rate $\kappa_j$ and detuning $\Delta_j$ appearing in Eq.~\eqref{QLEa} are given by $\Delta_{j}=\omega_{j}-\omega_{d}+\Delta_{\text{eff},j}$ and $\kappa_{j}=\kappa_{0}+\kappa_{\text{eff},j}$, where $\kappa_{0}$ is the intrinsic decay rate of the cavity mode.
The interference-induced terms $\kappa_{{\rm eff},j}$ and $\Delta_{{\rm eff},j}$ are defined in Eqs.~\eqref{eq:kappaeff_general} and~\eqref{eq:Deltaeff_general}, respectively.

For the coherent mean field generated by the cooling drive, the relevant propagation phase is the phase evaluated at the drive frequency.
We choose a reference drive frequency $\omega_{d,0}$ and define $\delta_d=\omega_{d,{\rm c}}-\omega_{d,0}$.
The corresponding accumulated phase is
\begin{equation}
\phi_{j,d}
=
\phi_{j,0}
+
\delta_d L/v_g ,
\label{eq:phi_drive}
\end{equation}
where $\phi_{j,0}$ is the phase at $\omega_{d,0}$, $L$ is the propagation distance between neighboring coupling points, and $v_g$ is the group velocity in the fiber. 
Accordingly, the phase-dependent quantities entering the mean-field equations are evaluated at $\phi_{j,d}$:
$E_{{\rm eff},j}=E\mathcal A_j(\phi_{j,d})$,
$\kappa_j=\kappa_0+\kappa_{{\rm eff},j}(\phi_{j,d})$, and
$\Delta_j=\omega_j-\omega_{d,{\rm c}}+\Delta_{{\rm eff},j}(\phi_{j,d})$.
We then linearize the dynamics by writing $\hat a_j
=
\alpha_j+\delta\hat a_j$, $\hat q
=
\bar q_c+\delta\hat q$, and $\hat p
=
\bar p_c+\delta\hat p$.
They satisfy
\begin{subequations}
\label{eq:mean_values_cooling}
\begin{align}
\alpha_j
&=\frac{E_{\text{eff},j}
}{\kappa_{j}+i\tilde\Delta_{j}},
\\
\bar q_c&=\frac{g|\alpha_j|^2}{\omega_m},
\\
\bar p_c&=0,
\end{align}
\end{subequations}
where the effective detuning including the radiation-pressure shift is
\begin{equation}
\tilde\Delta_{j}
=\Delta_{j}
-g\bar q_c
=\Delta_{j}
-\frac{g^2}{\omega_m}
|\alpha_j|^2.
\label{eq:Deltatilde_drive_phase}
\end{equation}
Equivalently, the mean intracavity photon number $n_j=|\alpha_j|^2$ is determined self-consistently by
\begin{equation}
n_j
\left[
\kappa_{j}^2
+\left(\Delta_{j}-
\frac{g^2}{\omega_m}n_j
\right)^2\right]
=\left|E_{\text{eff},j}
\right|^2.
\label{eq:cubic_meanfield_cooling}
\end{equation}
When more than one positive solution exists, the physical branch is selected by the dynamical stability condition discussed below.

We choose the reference drive frequency $\omega_{d,0}$ such that the effective red-sideband condition is satisfied at $\delta_d=0$,
\begin{equation}
\tilde\Delta_{j}(\phi_{j,0})
=\omega_j-\omega_{d,0}
+\Delta_{{\rm eff},j}(\phi_{j,0})
-g\bar q_{c,0}
=\omega_m.
\label{eq:red_sideband_condition}
\end{equation}
Here $\bar q_{c,0}$ denotes the steady-state displacement evaluated at $\delta_d=0$.
This condition differs from the usual single-point red-sideband condition because the optical resonance is additionally shifted by the phase-dependent self-interference term $\Delta_{{\rm eff},j}(\phi_{j,0})$.

We introduce the fluctuation-quadrature vector
\begin{equation}
\hat{\mathbf u}_{{\rm c},j}^{T}
=
\left(
\delta\hat x_j,
\delta\hat y_j,
\delta\hat q,
\delta\hat p
\right),
\label{eq:u_cooling}
\end{equation}
where
$\delta\hat x_j=(\delta\hat a_j+\delta\hat a_j^\dagger)/\sqrt{2}$
and
$\delta\hat y_j=i(\delta\hat a_j^\dagger-\delta\hat a_j)/\sqrt{2}$.
Defining
$G_j=\sqrt{2}g\alpha_j\equiv G_j^x+iG_j^y$,
the linearized fluctuation equations take the form
\begin{equation}
\dot{\hat{\mathbf u}}_{{\rm c},j}(t)
=
K_{{\rm c},j}\hat{\mathbf u}_{{\rm c},j}(t)
+
\hat{\mathbf v}_{{\rm c},j}(t),
\label{eq:linearQLE_drive_phase}
\end{equation}
with the drift matrix
\begin{equation}
K_{{\rm c},j}
=
\begin{pmatrix}
-\kappa_{j} & \tilde\Delta_{j} & -G_j^y & 0\\
-\tilde\Delta_{j} & -\kappa_{j} & G_j^x & 0\\
0 & 0 & 0 & \omega_m\\
G_j^x & G_j^y & -\omega_m & -\gamma_m
\end{pmatrix}.
\label{eq:K_cooling_drive_phase}
\end{equation}
Within the drive-phase approximation, all phase-dependent quantities are evaluated at $\phi_{j,d}$.

The corresponding noise vector is
\begin{equation}
\hat{\mathbf v}_{{\rm c},j}=\begin{pmatrix}
    \sqrt{2\kappa_{{\rm ex},j}}
\left[
{\rm Re}\mathcal A_j(\phi_{j,d})\hat x_j^{\rm in}
-
{\rm Im}\mathcal A_j(\phi_{j,d})\hat y_j^{\rm in}
\right]\\
\sqrt{2\kappa_{{\rm ex},j}}
\left[
{\rm Im}\mathcal A_j(\phi_{j,d})\hat x_j^{\rm in}
+
{\rm Re}\mathcal A_j(\phi_{j,d})\hat y_j^{\rm in}
\right]\\
    0\\
    \hat{\xi}\\
\end{pmatrix},
\label{vcj}
\end{equation}
where
$\hat x_j^{\rm in}
=(\hat a_j^{\rm in}+\hat a_j^{{\rm in},\dagger})/\sqrt{2}$
and
$\hat y_j^{\rm in}
=i(\hat a_j^{{\rm in},\dagger}-\hat a_j^{\rm in})/\sqrt{2}$.
To highlight the interference-induced optical response, we have considered the \emph{overcoupled regime} and neglected the intrinsic cavity loss, setting $\kappa_0\simeq0$~\cite{small1,small2}.
Because the linearized equations are driven by Gaussian Markovian noises, the stable steady state is Gaussian and is therefore fully characterized by its covariance matrix~\cite{Genes2008}, $\left(V_{{\rm c},j}\right)_{kl}(t)=\left\langle
\left(\hat{\mathbf u}_{{\rm c},j}\right)_k(t)
\left(\hat{\mathbf u}_{{\rm c},j}\right)_l(t)
+
\left(\hat{\mathbf u}_{{\rm c},j}\right)_l(t)
\left(\hat{\mathbf u}_{{\rm c},j}\right)_k(t)
\right\rangle/2$. 
Its time evolution obeys the differential Lyapunov equation
\begin{equation}
\frac{dV_{{\rm c},j}}{dt}
=K_{{\rm c},j}V_{{\rm c},j}
+V_{{\rm c},j}K_{{\rm c},j}^{T}
+D_{{\rm c},j},
\label{le}
\end{equation}
where $D_{{\rm c},j}$ is the diffusion matrix defined as $\left\langle
\left(\hat{\mathbf v}_{{\rm c},j}\right)_k(t)
\left(\hat{\mathbf v}_{{\rm c},j}\right)_l(t')
+
\left(\hat{\mathbf v}_{{\rm c},j}\right)_l(t')
\left(\hat{\mathbf v}_{{\rm c},j}\right)_k(t)
\right\rangle/2
=
\left(D_{{\rm c},j}\right)_{kl}\delta(t-t')$.
Substituting Eq.~\eqref{vcj} into this definition gives
$D_{{\rm c},j}=\operatorname{diag}
\left[
F_j(\phi_{j,d}),
F_j(\phi_{j,d}),
0,
\gamma_m(2n_m+1)
\right]$.
The steady-state covariance matrix is then obtained from the Lyapunov equation
\begin{equation}
K_{{\rm c},j}V_{{\rm c},j}
+
V_{{\rm c},j}K_{{\rm c},j}^{T}
+
D_{{\rm c},j}
=
0.
\label{eq:Lyapunov_cooling_drive_phase}
\end{equation}

We next examine the dynamical stability of the linearized cooling system.
The steady-state solution is retained only when all eigenvalues of the drift matrix $K_{{\rm c},j}$ have negative real parts, namely
\begin{equation}
\max_{\ell}
\left[
\operatorname{Re}
\{
\lambda_\ell(K_{{\rm c},j})
\}
\right]
<0,
\label{stability_cooling}
\end{equation}
where $\lambda_\ell(K_{{\rm c},j})$ denotes the $\ell$th eigenvalue of $K_{{\rm c},j}$. 
The steady-state mechanical phonon number is calculated from the mechanical block of the covariance matrix,
\begin{equation}
\bar n_j
=
\frac{1}{2}
\left[
(V_{{\rm c},j})_{33}
+
(V_{{\rm c},j})_{44}
-
1
\right].
\label{eq:nbar_cooling_drive_phase}
\end{equation}

Strictly speaking, the propagation phase relevant for the optical fluctuations should be associated with the scattered optical field rather than with the coherent drive.
In the rotating frame of the drive, a fluctuation component at Fourier frequency $\nu$ corresponds to the laboratory-frame optical frequency $\omega_{d,{\rm c}}+\nu$. Its propagation phase is therefore
\begin{equation}
\phi_j(\nu)=\phi_{j,d}+\nu L/v_g.
\label{eq:phi_scattered}
\end{equation}
In a fully frequency-dependent treatment, the optical response should therefore be evaluated with
$\kappa_j(\nu)=\kappa_0+\kappa_{{\rm eff},j}[\phi_j(\nu)]$,
$\Delta_{{\rm eff},j}[\phi_j(\nu)]$, and
$F_j[\phi_j(\nu)]$,
instead of their drive-phase values at $\phi_{j,d}$.
The corresponding steady-state covariance matrix can then be obtained from the frequency-domain spectrum,
\begin{equation}
\begin{split}
V_{{\rm c},j}^{\rm fd}
=&
\int_{-\infty}^{+\infty}
\frac{d\nu}{2\pi}
\left[
-i\nu I
-
K_{{\rm c},j}(\nu)
\right]^{-1}
D_{{\rm c},j}(\nu)
\\
&\times
\left(
\left[
-i\nu I
-
K_{{\rm c},j}(\nu)
\right]^{-1}
\right)^{\dagger}.
\label{eq:V_frequency_dependent}
\end{split}
\end{equation}
In the main text, we use the drive-phase approximation to elucidate the underlying physics, namely
\begin{equation}
\phi_j(\nu)\simeq \phi_{j,d}.
\label{eq:drive_phase_approx}
\end{equation}
Under this approximation, the phase-dependent quantities entering the cooling dynamics are evaluated at the coherent driving frequency. 
The influence of using the scattered-field phase instead is examined in Appendix~\ref{app:cooling_phase_check}, where we compare the drive-phase result with the frequency-dependent calculation. 
The comparison shows that the cooling window and the contour $\bar n_j=1$ remain qualitatively unchanged.

Throughout this work, the system is
treated within the short-delay Markov approximation. The
maximum propagation delay across the $N$ coupling points is
$\tau_{\rm max}=(N-1)L/v_g$. For $N=4$ and the parameters used in our work,
$\kappa_{{\rm ex},j}\tau_{\rm max}\simeq4.1\times10^{-2}\ll1$.
Therefore, the non-Markovian
retardation can be safely neglected.

\section{Phase-dependent and nonreciprocal phonon cooling}
\label{sec:cooling_results}

We first discuss phase-dependent cooling in the absence of rotation, $\Omega=0$. 
In this case, the CW and CCW optical modes are degenerate. 
The multi-point self-interference therefore produces the same phase-dependent effective linewidth and detuning for left and right incidence, leading to a reciprocal cooling response.

\begin{figure}[t]
\centering
\includegraphics[width=0.37\textwidth]{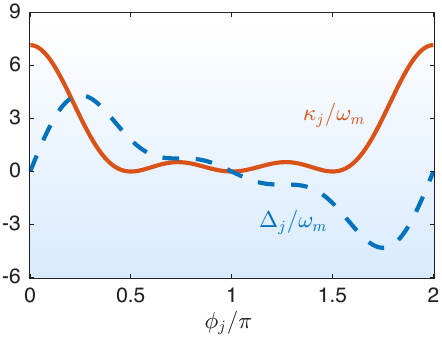}
\caption{Phase dependence of the effective optical linewidth $\kappa_j/\omega_m$ (solid) and detuning $\Delta_j/\omega_m$ (dashed) for $N=4$ in the overcoupled limit $\kappa_0\simeq0$. 
Multi-point self-interference strongly modulates the linewidth, allowing the system to be tuned between the resolved- and unresolved-sideband regimes. The parameters are
$\kappa_{{\rm ex},j}\simeq38\,{\rm MHz}$ and $\omega_m\simeq85\,{\rm MHz}$.}
\label{fig:kappa_Delta}
\end{figure}

For $N=4$, the effective optical linewidth is strongly modulated by the propagation phase through multi-point self-interference.
Figure~\ref{fig:kappa_Delta} shows $\kappa_j/\omega_m$ as a function of the phase $\phi_j$ in the overcoupled limit $\kappa_0\simeq0$.
This modulation originates from the cosine interference term in Eq.~\eqref{eq:kappaeff_general}. 
The system can therefore be tuned between the resolved-sideband regime, $\kappa_j<\omega_m$, and the unresolved-sideband regime, $\kappa_j\gtrsim\omega_m$, simply by changing the phase accumulated between neighboring coupling points. 
This phase-dependent linewidth is essential for the cooling behavior discussed below, because sideband cooling is highly sensitive to the ratio $\kappa_j/\omega_m$.

\begin{figure}[t]
\centering
\includegraphics[width=0.42\textwidth]{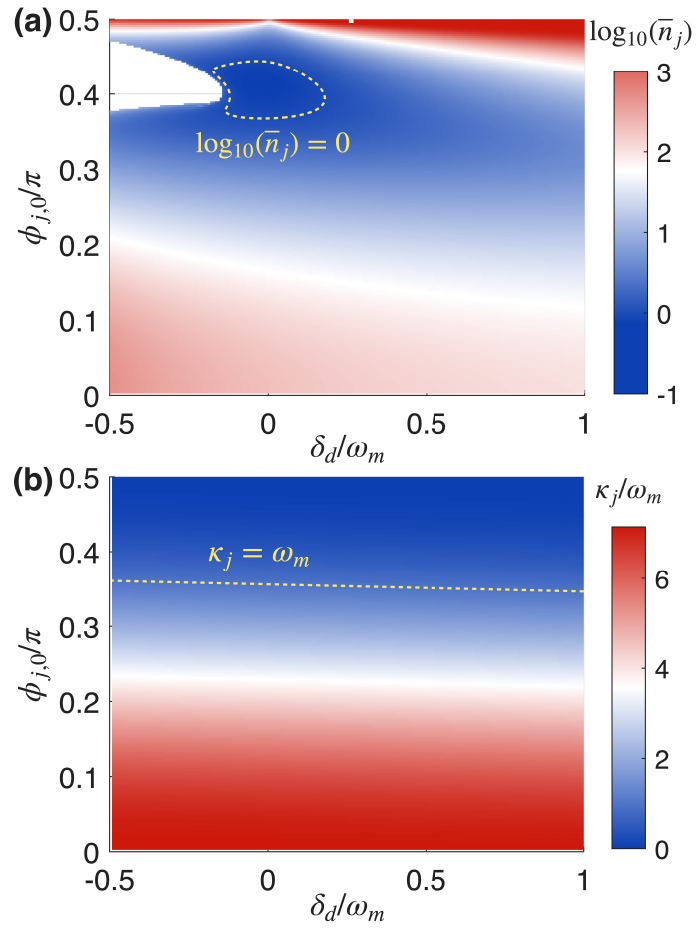}
\caption{
Phase-dependent cooling for a static cavity. 
(a) Logarithm of the steady-state phonon number, $\log_{10}(\bar n_j)$, and 
(b) effective linewidth, $\kappa_j/\omega_m$, as functions of the driving detuning $\delta_d$ and the reference phase $\phi_{j,0}$. 
The yellow dashed contour in (a) and dashed line in (b) mark the $\bar n_j=1$ threshold and the boundary $\kappa_j=\omega_m$ between the resolved- and unresolved-sideband regimes, respectively. 
The parameters are
$\kappa_{{\rm ex},j}\simeq38\,{\rm MHz}$,
$\omega_m\simeq85\,{\rm MHz}$,
$\gamma_m\simeq3\,{\rm kHz}$,
$T\simeq3\,{\rm K}$,
$g\simeq67\,{\rm Hz}$,
$P\simeq5.4\,{\rm W}$,
$L\simeq4\times10^{-3}\,{\rm m}$,
and $v_g\simeq1.1\times10^{7}\,{\rm m/s}$.
}
\label{fig:PhaseDep}
\end{figure}

Figure~\ref{fig:PhaseDep} shows the steady-state phonon number and the effective linewidth in the $(\delta_d/\omega_m,\phi_{j,0}/\pi)$ plane. 
The reference phase $\phi_{j,0}$ fixes the self-interference condition at the reference drive frequency, while $\delta_d$ changes both the optical detuning and the propagation phase according to Eq.~\eqref{eq:phi_drive}. Efficient cooling is achieved only within a finite region where two requirements are simultaneously satisfied. 
First, the effective detuning must remain close to the red-sideband condition, $\tilde\Delta_j\simeq\omega_m$.
Second, the optical linewidth must be small enough to provide sideband resolution, but not so small that the optical leakage channel becomes ineffective. 
Thus, neither an extremely large nor an extremely small effective linewidth is optimal for cooling.

The validity of the drive-phase approximation is further examined in Appendix~\ref{app:cooling_phase_check}, where the calculation is repeated using the frequency-dependent scattered-field phase $\phi_j(\nu)$. 
As shown in Fig.~\ref{fig:app_cooling_phase_check}, the resulting $\bar n_j^{\rm fd}=1$ boundary closely follows the $\bar n_j=1$ boundary obtained from the drive-phase calculation. 
Thus, the drive-phase approximation captures the relevant phase-dependent cooling behavior in the parameter regime considered here.

We now consider the effect of cavity rotation. 
The Sagnac--Fizeau shift separates the CW and CCW resonance frequencies according to Eq.~\eqref{eq:cw_ccw_freq}. 
Consequently, the two counterpropagating modes experience different effective detunings, and their corresponding resonance conditions occur at different reference phases. 
Because the multi-point interference makes the effective linewidth strongly phase dependent, the two directions can enter different sideband-resolution regimes: one direction may satisfy $\kappa_j<\omega_m$, while the opposite direction remains in the unresolved-sideband regime with $\kappa_j\gtrsim\omega_m$.

\begin{figure}[t]
\centering
\includegraphics[width=0.42\textwidth]{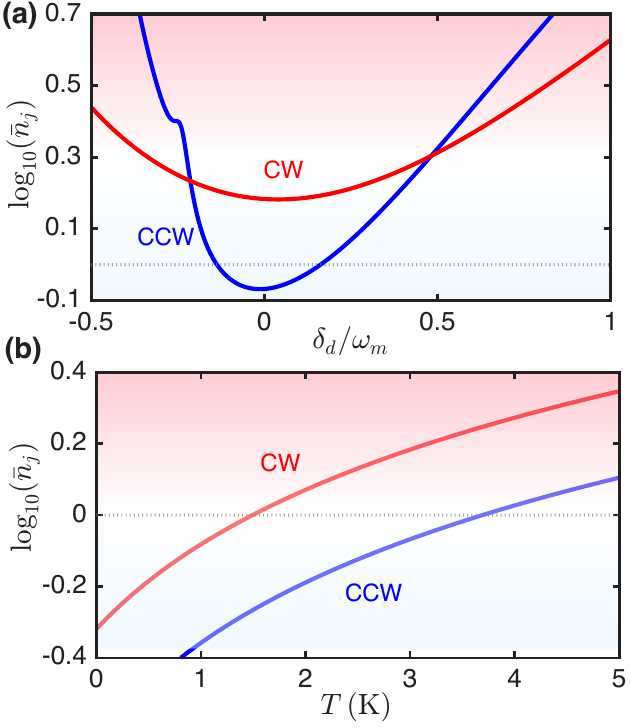}
\caption{
Nonreciprocal phonon cooling for the two opposite driving directions.
(a) Logarithm of the steady-state phonon number, $\log_{10}(\bar n_j)$, as a function of the driving detuning $\delta_d$ at $T\simeq3\,{\rm K}$.
(b) $\log_{10}(\bar n_j)$ as a function of the temperature $T$ at the reference red-sideband point $\delta_d\simeq0$.
The reference phases are chosen as $\phi_{\rm cw,0}\simeq0.34\pi$ and $\phi_{\rm ccw,0}\simeq0.38\pi$, corresponding to a phase difference $\delta\phi_0\simeq0.04\pi$ between the two counterpropagating modes. For the spinning cavity considered here, we take $|\Delta_F|=2.08\omega_m$, corresponding to $\Omega\approx30\,{\rm kHz}$. The other parameters are the same as those in Fig.~\ref{fig:PhaseDep}.
}
\label{fig:cooling_cw_ccw}
\end{figure}

This direction-dependent optical response leads to nonreciprocal phonon cooling, as shown in Fig.~\ref{fig:cooling_cw_ccw}.
For the chosen parameters, the CCW mode enters the resolved-sideband regime and cools the mechanical mode close to its ground state, whereas the CW mode has a larger effective linewidth and results in higher phonon numbers. 
Figure~\ref{fig:cooling_cw_ccw}(a) shows $\log_{10}(\bar n_j)$ as a function of the driving detuning $\delta_d$, while Fig.~\ref{fig:cooling_cw_ccw}(b) shows its temperature dependence at the reference red-sideband point $\delta_d\simeq0$.
Thus, the nonreciprocity arises from 
the interplay between the rotation-induced Sagnac--Fizeau splitting and the phase-dependent self-interference of the multi-point coupling.
As a result, cooling close to the ground-state regime can be achieved for one input direction, while the opposite direction remains much less efficiently cooled.

\section{Two-tone driving and mechanical squeezing}
\label{sec:squeezing_theory}

We now use the same phase-dependent self-interference mechanism
to engineer steady-state mechanical squeezing. Two-tone or
modulated driving provides a well-established route to
reservoir-engineered mechanical squeezing
~\cite{Mari2009,Kronwald2013,Wollman2015,Lei2016}.
In contrast to the cooling configuration in Sec.~\ref{sec:cooling_theory}, where a single red-detuned tone is applied,  the selected optical mode $j$ is now driven simultaneously by two coherent tones. 
The two tones are placed close to the red and blue mechanical sidebands, respectively.
When the effective red-sideband coupling exceeds the blue-sideband one, the driven cavity acts as an engineered squeezed reservoir for the mechanical resonator~\cite{Kronwald2013}.

We choose a reference center frequency $\omega_{d,0}$ for the two-tone drive.
With an additional detuning $\delta_d$, the two-tone center frequency is
\begin{equation}
\omega_{d,\rm s}=\omega_{d,0}+\delta_d.
\label{eq:sq_center_freq}
\end{equation}
The red- and blue-sideband drive frequencies are then
\begin{equation}
\omega_{d,-}=\omega_{d,\rm s}-\omega_m,
\quad
\omega_{d,+}=\omega_{d,\rm s}+\omega_m,
\label{eq:sq_twotone_freq}
\end{equation}
where the subscripts $-$ and $+$ denote the red- and blue-sideband tones, respectively. 
The driving Hamiltonian reads
\begin{equation}
H_{d,{\rm s}}
=i\hbar\left[E_{{\rm eff},j}^{(-)}
\hat a_j^\dagger e^{-i\omega_{d,-}t}
+E_{{\rm eff},j}^{(+)}
\hat a_j^\dagger e^{-i\omega_{d,+}t}
-{\rm H.c.}\right],
\label{eq:Hdrive_squeezing}
\end{equation}
where the subscript ``s'' denotes the squeezing configuration.

The reference two-tone center frequency $\omega_{d,0}$ is chosen to be resonant with the interference-shifted optical resonance, namely
\begin{equation}
\tilde{\Delta}_{j,\rm s}(\phi_{j,0})
=\omega_j-\omega_{d,0}
+\Delta_{{\rm eff},j}(\phi_{j,0})-g\bar q_{s,0}=0,
\label{eq:sq_center_condition}
\end{equation}
where at the reference point $\delta_d=0$, this
displacement is $\bar q_{s,0}=g\left(|\alpha_{j,-}|^2+|\alpha_{j,+}|^2\right)/\omega_m$ with $\alpha_{j,-}$ and $\alpha_{j,+}$ being defined later.
This condition differs from the cooling reference condition in Eq.~\eqref{eq:red_sideband_condition}. 
In the cooling case, $\delta_d=0$ corresponds to the effective red-sideband condition, $\tilde\Delta_j=\omega_m$. 
In the present squeezing case, $\delta_d=0$ corresponds instead to the resonance condition of the two-tone center, $\tilde\Delta_{j,\rm s}=0$.

The propagation phase of the two-tone center frequency is written in the same form as in the cooling section,
\begin{equation}
\phi_{j,\rm s}=\phi_{j,0}+
\delta_d L/v_g.
\label{eq:sq_phi_center}
\end{equation}
Since the red and blue tones have different optical frequencies, they acquire slightly different propagation phases,
\begin{equation}
\phi_{j,-}
=\phi_{j,\rm s}-\delta\phi_m,
\quad
\phi_{j,+}=\phi_{j,\rm s}+
\delta\phi_m,
\label{eq:sq_phi_pm}
\end{equation}
with $\delta\phi_m=\omega_m L/v_g$.
Thus, the red and blue coherent tones experience the same type of multi-point self-interference introduced in Sec.~\ref{sec:general_model}, but evaluated at their respective phases $\phi_{j,\pm}$.

Using Eq.~\eqref{eq:sq_center_condition}, the effective detuning of the two-tone center at the detuning $\delta_d$ is
\begin{equation}
\begin{split}
\tilde{\Delta}_{j,\rm s}
=&-\delta_d-
\Delta_{{\rm eff},j}(\phi_{j,0})
+\Delta_{{\rm eff},j}(\phi_{j,\rm s})\\
&-g\left[
\bar q_{{\rm s}}(\delta_d)
-\bar q_{{\rm s},0}\right].
\label{eq:sq_Delta_center}
\end{split}
\end{equation}
Similarly, the effective detunings associated with  the coherent red and blue tones are
\begin{subequations}
\label{eq:sq_Delta_pm}
\begin{align}
\tilde{\Delta}_{j,-}
=&-\delta_d
-\Delta_{{\rm eff},j}(\phi_{j,0})
+\Delta_{{\rm eff},j}(\phi_{j,-})
+\omega_m
\nonumber\\
&-g\left[
\bar q_{{\rm s}}(\delta_d)
-\bar q_{{\rm s},0}
\right],\\
\tilde{\Delta}_{j,+}
=&-\delta_d
-\Delta_{{\rm eff},j}(\phi_{j,0})
+\Delta_{{\rm eff},j}(\phi_{j,+})
-\omega_m
\nonumber\\
&-g\left[
\bar q_{{\rm s}}(\delta_d)
-\bar q_{{\rm s},0}
\right].
\end{align}
\end{subequations}
At $\delta_d=0$, the red and blue tones are therefore placed close to the effective red and blue mechanical sidebands, respectively.

The effective driving amplitudes of the two tones are
\begin{equation}
\label{eq:sq_Eeff_pm}
E_{{\rm eff},j}^{(-)}
=E_-
\mathcal A_j(\phi_{j,-}),
\quad
E_{{\rm eff},j}^{(+)}
=E_+
\mathcal A_j(\phi_{j,+}),
\end{equation}
where $\mathcal A_j(\phi)$ is the multi-point interference amplitude defined in Eq.~\eqref{eq:A_phi_general}. 
The corresponding effective linewidths are
\begin{equation}
\label{eq:sq_kappa_pm}
\kappa_{j,-}
=\kappa_0+\kappa_{{\rm eff},j}(\phi_{j,-}),
\quad
\kappa_{j,+}
=\kappa_0+\kappa_{{\rm eff},j}(\phi_{j,+}).
\end{equation}
Once again, we consider the overcoupled limit $\kappa_0\simeq0$. Therefore, the coherent intracavity amplitudes generated by the two tones are
\begin{equation}
\label{eq:sq_alpha_pm}
\alpha_{j,-}
=\frac{
E_{{\rm eff},j}^{(-)}
}{\kappa_{j,-}
+i\tilde{\Delta}_{j,-}
},
\quad
\alpha_{j,+}
=\frac{
E_{{\rm eff},j}^{(+)}
}{\kappa_{j,+}
+i\tilde{\Delta}_{j,+}}.
\end{equation}
The effective red- and blue-sideband optomechanical couplings are then $G_{j,-}=\sqrt{2}g|\alpha_{j,-}|$ and $G_{j,+}=\sqrt{2}g|\alpha_{j,+}|$, respectively.
The phases of the coherent amplitudes determine the orientation of the squeezed mechanical quadrature. Since we focus on the squeezing strength, we choose the quadrature axes such that $G_{j,-}$ and $G_{j,+}$ are real and nonnegative.

The optical fluctuation mode used in the covariance-matrix dynamics is defined in the rotating frame of the two-tone center frequency. 
Accordingly, the optical linewidth and noise strength entering the fluctuation dynamics are evaluated at the center phase,
\begin{equation}
\kappa_{j,\rm s}
=\kappa_{{\rm eff},j}(\phi_{j,\rm s}),
\quad
F_{j,\rm s}=F_j(\phi_{j,\rm s}).
\label{eq:sq_kappa_F_center}
\end{equation}
Here $F_{j,\rm s}$ denotes the optical diffusion strength associated with the external waveguide channel. 
As in the cooling analysis, the phase accumulated by a fluctuation at Fourier frequency $\nu$ is, strictly speaking, frequency dependent, $\phi_j(\nu)=\phi_{j,s}+\nu L/v_g$. 
Here, we employ the center-phase approximation $\phi_j(\nu)\simeq\phi_{j,s}$, whose validity for the squeezing calculation is examined in Appendix~\ref{app:squeezing_phase_check}.

After moving to the mechanical rotating frame and applying the rotating-wave approximation (RWA), the linearized interaction Hamiltonian becomes
\begin{equation}
H_{{\rm lin},j}^{\rm RWA}
=
-\frac{\hbar G_{j,-}}{2}
\delta\hat a_j^\dagger \delta\hat b-\frac{\hbar G_{j,+}}{2}
\delta\hat a_j^\dagger \delta\hat b^\dagger+{\rm H.c.},
\label{eq:Hlin_squeezing_RWA}
\end{equation}
where $\delta\hat b
=(\delta\hat Q+i\delta\hat P)/\sqrt{2}$
is the mechanical fluctuation operator in the rotating frame. 
For $G_{j,+}<G_{j,-}$, one can introduce a mechanical Bogoliubov mode, $\hat\beta_j
=\delta\hat b\cosh r_j
+\delta\hat b^\dagger\sinh r_j$, with $\tanh r_j=G_{j,+}/G_{j,-}$, as
\begin{equation}
H_{{\rm lin},j}^{\rm RWA}
=-\hbar \mathcal G_j
\delta\hat a_j^\dagger \hat\beta_j
+{\rm H.c.},
\label{eq:Hbeta_squeezing}
\end{equation}
where $\mathcal G_j
=\frac{1}{2}\sqrt{G_{j,-}^2-G_{j,+}^2}$.
The optical dissipation then cools the Bogoliubov mode $\hat\beta_j$. 
Since the vacuum of $\hat\beta_j$ corresponds to a squeezed state of the original mechanical mode, this dissipative process generates steady-state mechanical squeezing~\cite{Kronwald2013}.

To quantify the mechanical squeezing, we define the fluctuation vector
\begin{equation}
\hat{\mathbf u}_{{\rm s},j}^{T}
=
\left(
\delta\hat X_j,
\delta\hat Y_j,
\delta\hat Q,
\delta\hat P
\right),
\label{eq:u_squeezing}
\end{equation}
where $\delta\hat X_j$ and $\delta\hat Y_j$ are the optical quadratures in the rotating frame of the two-tone center frequency, and $\delta\hat Q$ and $\delta\hat P$ are the mechanical quadratures in the mechanical rotating frame.
We further define $G_{j,\Sigma}
=G_{j,-}+G_{j,+}$ and $G_{j,D}=
G_{j,-}-G_{j,+}$.
The linearized RWA Langevin equation can then be written as
\begin{equation}
\dot{\hat{\mathbf u}}_{{\rm s},j}(t)
=
K_{{\rm s},j}\hat{\mathbf u}_{{\rm s},j}(t)
+
\hat{\mathbf v}_{{\rm s},j}(t),
\label{eq:QLE_squeezing}
\end{equation}
with the drift matrix
\begin{equation}
K_{{\rm s},j}
=
\begin{pmatrix}
-\kappa_{j,s} & \tilde\Delta_{j,s} & 0 & -G_{j,D}/2\\
-\tilde\Delta_{j,s} & -\kappa_{j,s} & G_{j,\Sigma}/2 & 0\\
0 & -G_{j,D}/2 & -\gamma_m/2 & 0\\
G_{j,\Sigma}/2 & 0 & 0 & -\gamma_m/2
\end{pmatrix}.
\label{eq:K_squeezing}
\end{equation}
The corresponding noise vector is
\begin{equation}
\hat{\mathbf v}_{{\rm s},j}
=
\begin{pmatrix}
\sqrt{2\kappa_{{\rm ex},j}}
\left[
{\rm Re}\mathcal A_j(\phi_{j,s})\hat X_j^{\rm in}
-
{\rm Im}\mathcal A_j(\phi_{j,s})\hat Y_j^{\rm in}
\right]
\\
\sqrt{2\kappa_{{\rm ex},j}}
\left[
{\rm Im}\mathcal A_j(\phi_{j,s})\hat X_j^{\rm in}
+
{\rm Re}\mathcal A_j(\phi_{j,s})\hat Y_j^{\rm in}
\right]
\\
\hat\xi_Q
\\
\hat\xi_P
\end{pmatrix}.
\label{eq:squeezing_noise_vector}
\end{equation}
Here $\hat\xi_Q$ and $\hat\xi_P$ denote the effective mechanical
noise components in the rotating frame. Before applying the RWA,
they are related to the Brownian force $\hat\xi$ introduced in
Eq.~\eqref{QLEc} by $\hat\xi_Q(t)
=-\hat\xi(t)\sin(\omega_m t)$ and $\hat\xi_P(t)=\hat\xi(t)\cos(\omega_m t)$.
After period averaging within the RWA, their symmetrized
correlations become $\frac{1}{2}
\left\langle
\left\{
\hat\xi_\mu(t),\hat\xi_\nu(t')
\right\}
\right\rangle
=
\frac{\gamma_m}{2}(2n_m+1)
\delta_{\mu\nu}\delta(t-t')$ with $\mu,\nu=Q,P$.
The diffusion matrix is $D_{{\rm s},j}
={\rm diag}\left[F_{j,s},
F_{j,s},\gamma_m(n_m+\frac{1}{2}),\gamma_m(n_m+\frac{1}{2})\right]$.
Here $F_{j,s}$ is evaluated at the two-tone center phase, while the mechanical thermal noise appears symmetrically in the rotating mechanical quadratures.

The steady-state covariance matrix $V_{{\rm s},j}$ is obtained from the Lyapunov equation
\begin{equation}
K_{{\rm s},j}V_{{\rm s},j}
+V_{{\rm s},j}K_{{\rm s},j}^{T}
+D_{{\rm s},j}=0.
\label{eq:Lyapunov_squeezing}
\end{equation}
Only dynamically stable points satisfying 
\begin{equation}
\max_{\ell}
\left[{\rm Re}\{\lambda_\ell(K_{{\rm s},j})\}\right]<0
\label{eq:stability_squeezing}
\end{equation}
are retained. 
In addition, stable dissipative squeezing requires $G_{j,+}<G_{j,-}$.

The minimum mechanical quadrature variance is obtained from the mechanical block of the covariance matrix. 
The explicit expression is
\begin{equation}
\begin{split}
V_{\min,j}
=&\frac{(V_{{\rm s},j})_{33}
+(V_{{\rm s},j})_{44}}{2}
\\
&-\sqrt{\left[\frac{(V_{{\rm s},j})_{33}
-(V_{{\rm s},j})_{44}}{2}
\right]^2+(V_{{\rm s},j})_{34}^{2}}.
\end{split}
\label{eq:Vmin_squeezing}
\end{equation}
This expression is obtained by diagonalizing the mechanical covariance block; see Appendix~\ref{app:Vmin}.
Since the vacuum variance is $1/2$ here, the mechanical squeezing level in dB is defined as
\begin{equation}
\mathcal S_j=-10\log_{10}
\left(2V_{\min,j}\right).
\label{eq:Squeezing_dB}
\end{equation}
Thus, $\mathcal S_j>0$ indicates squeezing below the zero-point fluctuation level, while $\mathcal S_j>3\,{\rm dB}$ corresponds to squeezing beyond the 3-${\rm dB}$ threshold.

The RWA formulation above provides a transparent engineering picture in terms of the red- and blue-sideband couplings $G_{j,-}$ and $G_{j,+}$. 
However, because the phase-dependent optical linewidth can become comparable to the mechanical frequency in some regions of parameter space, the counter-rotating terms may introduce quantitative corrections. 
In Appendix~\ref{app:rwa_check}, we compare the RWA squeezing map with a time-periodic covariance calculation including the counter-rotating terms. 
The comparison shows that the main region of strong squeezing remains robust beyond the RWA, although the RWA tends to overestimate the extent of the squeezed region.

\section{Phase-dependent and nonreciprocal mechanical squeezing}
\label{sec:squeezing_results}

We now show the mechanical squeezing results obtained from the two-tone configuration described above. 
For each pair $(\delta_d,\phi_{j,0})$, the red-tone input power is fixed, while the blue-tone input amplitude is optimized. 
Equivalently, we scan the input ratio $E_+/E_-$ and select the stable solution with $G_{j,+}<G_{j,-}$ that gives the largest mechanical squeezing $\mathcal S_j$. 
The optimal value of $G_{j,+}/G_{j,-}$ is then determined by the phase-dependent mean fields in Eq.~\eqref{eq:sq_alpha_pm}, rather than imposed as an independent control parameter.

\begin{figure}[t]
\centering
\includegraphics[width=0.45\textwidth]{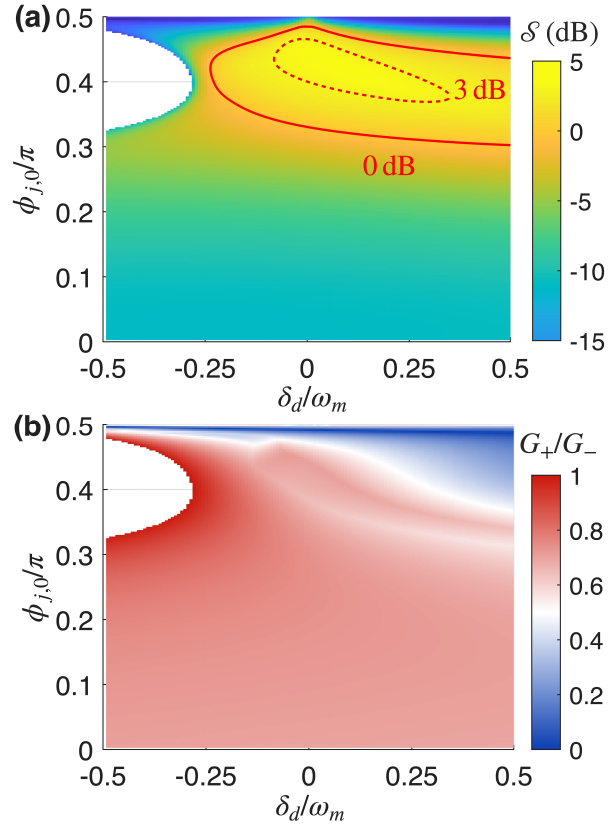}
\caption{
Phase-dependent mechanical squeezing for a static cavity. 
(a) Optimized mechanical squeezing $\mathcal S_j$ and 
(b) the corresponding optimal two-tone coupling ratio $G_{j,+}/G_{j,-}$ as functions of the driving detuning $\delta_d$ and the reference phase $\phi_{j,0}$. 
The red solid and dashed contours in (a) indicate $\mathcal S_j=0\,{\rm dB}$ and $\mathcal S_j=3\,{\rm dB}$, respectively.  
Only stable parameter points satisfying Eq.~\eqref{eq:stability_squeezing} and $G_{j,+}<G_{j,-}$ are retained. 
The red-tone input power is $P_-\simeq3.4\,{\rm W}$, the mechanical damping rate is $\gamma_m\simeq2\,{\rm kHz}$, and the temperature is $T\simeq0.1\,{\rm K}$. 
The other parameters are the same as those in Fig.~\ref{fig:PhaseDep}.
}
\label{fig:squeezing}
\end{figure}

Figure~\ref{fig:squeezing} shows the optimized mechanical squeezing and the corresponding optimal coupling ratio $G_{j,+}/G_{j,-}$ in the $(\delta_d/\omega_m,\phi_{j,0}/\pi)$ plane. 
Strong squeezing appears only within finite parameter regions. 
This behavior is a direct consequence of the multi-point self-interference: changing $\phi_{j,0}$ and $\delta_d$ modifies the effective driving amplitudes, effective detunings, and linewidths of the two tones. 
The requirement differs from cooling, which is optimized near an effective red sideband. 
For squeezing, the two-tone center should remain close to the interference-shifted optical resonance, while the blue-sideband coupling must be large enough to generate squeezing but smaller than the red-sideband coupling.
Therefore, the optimal squeezing results from a balance between sideband resolution, effective driving strength, and coupling ratio.

\begin{figure}[t]
\centering
\includegraphics[width=0.4\textwidth]{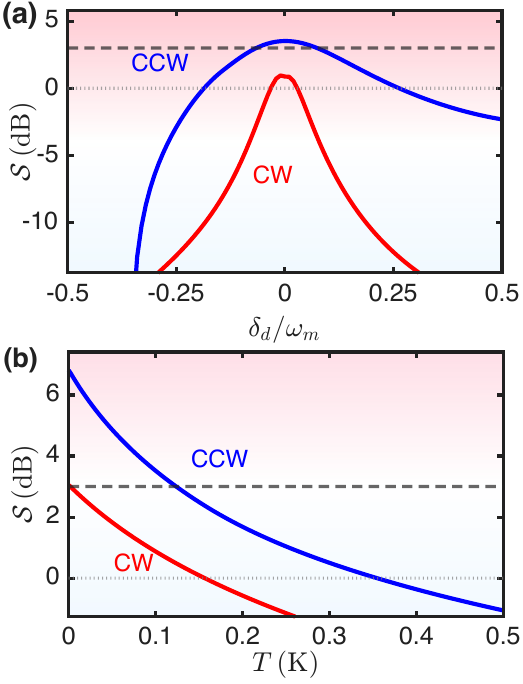}
\caption{
Nonreciprocal mechanical squeezing for the two opposite driving directions. 
(a) Optimized mechanical squeezing $\mathcal S_j$ as a function of the driving detuning $\delta_d$ at $T\simeq0.1\,{\rm K}$. 
(b) $\mathcal S_j$ as a function of the temperature $T$ at the two-tone-center resonance, $\delta_d\simeq0$. 
For each value of $\delta_d$ or $T$, the two-tone driving amplitudes are optimized, and the stable solution that gives the largest $\mathcal S_j$ is selected. 
The reference phases are chosen as $\phi_{\rm ccw,0}\simeq0.45\pi$ and $\phi_{\rm cw,0}\simeq0.48\pi$, corresponding to a phase difference $\delta\phi_0\simeq0.03\pi$ between the two counterpropagating modes. 
For the spinning case considered here, we take $|\Delta_F|=1.56\omega_m$, corresponding to $\Omega\approx22.5\,{\rm kHz}$. 
The other parameters are the same as those in Fig.~\ref{fig:squeezing}.
}
\label{fig:squeezing_cw_ccw}
\end{figure}

The same mechanism can be made nonreciprocal by switching on the cavity rotation. 
The Sagnac--Fizeau shift separates the CW and CCW optical resonance frequencies, leading to different effective detunings and shifting the optimal interference conditions for the two counterpropagating modes.
Because the two-tone couplings $G_{j,-}$ and $G_{j,+}$ depend sensitively on the phase-dependent effective driving amplitudes, effective detunings and linewidths, the squeezing condition can be satisfied for one propagation direction but not for the other.

This direction-dependent behavior is shown in Fig.~\ref{fig:squeezing_cw_ccw}. 
For each value of $\delta_d$ or $T$, we optimize the two-tone amplitudes and retain the stable solution with the largest squeezing. 
In the chosen parameter regime, mechanical squeezing is clearly visible for one driving direction, whereas the opposite direction gives a much weaker response.
Thus, the same self-interference mechanism responsible for phase-dependent cooling also enables nonreciprocal mechanical squeezing.

\section{Conclusion}
\label{sec:conclusion}

In conclusion, we have proposed a giant optomechanical system in which a spinning whispering-gallery-mode resonator is coupled to a meandering waveguide at multiple spatially separated points. 
The multi-point coupling structure leads to self-interference that simultaneously modifies the effective coherent drive, optical linewidth, and frequency shift. 
For a static cavity, these quantities can be controlled through the propagation phase, which provides a flexible way to engineer the optical reservoir and the sideband resolution. 
We have shown that this mechanism enables phase-dependent phonon cooling under a single red-detuned drive and steady-state mechanical squeezing under proper two-tone driving. 
When the cavity rotates, the Sagnac--Fizeau splitting causes resonant excitation of the CW and CCW modes to occur at different drive frequencies, thereby making the two modes sample different self-interference conditions. As a result, this interplay renders both the cooling and squeezing processes direction dependent. 
These results demonstrate that the combination of cavity rotation and multi-point self-interference provides a unified route to phase-controlled and nonreciprocal reservoir engineering in optomechanical systems.

The giant-cavity architecture considered here also opens several avenues for future research. 
Beyond the Gaussian cooling and squeezing processes investigated in this work, the strongly direction-dependent optical response could be exploited to realize more nonreciprocal quantum effects, including photon blockade, optomechanical entanglement, and quantum state transfer. 
Introducing optical or mechanical nonlinearities may further enable the selective preparation of nonclassical or even non-Gaussian states. 
In addition, the phases associated with the multi-point coupling points provide extra degrees of freedom for controlling the generation, storage, and transfer of quantum states between different optical and mechanical subsystems. 
Exploring the combined effects of the Sagnac--Fizeau shift, multi-point self-interference, and quantum nonlinearity may therefore lead to new promising schemes for directional quantum state engineering and nonreciprocal quantum information processing.

\begin{acknowledgments}
We thank Dr.\ Lei Du, Prof.\ Hui Jing, Prof.\ Oriol Romero-Isart for their valuable support.
This research has been supported by the European Research Council (ERC) under the grant agreement No. [951234] (Q-Xtreme ERC-2020-SyG). 
\end{acknowledgments}

\appendix

\section{Check of the drive-phase approximation for mechanical cooling}
\label{app:cooling_phase_check}

For the cooling calculation presented in the main text, the phase-dependent optical response is evaluated at the coherent driving frequency, i.e., the drive-phase approximation $\phi_j(\nu)\simeq\phi_{j,d}$ is used. 
Here we check the effectiveness of this approximation by replacing the drive phase with the frequency-dependent scattered-field phase
$\phi_j(\nu)=\phi_{j,d}+\nu L/v_g$, as introduced in Sec.~\ref{sec:cooling_theory}.
With this replacement, the optical linewidth, interference-induced frequency shift, and optical diffusion strength become frequency dependent, and the steady-state covariance matrix is obtained from the frequency-domain noise spectrum,
\begin{equation}
\begin{split}
V_{{\rm c},j}^{\rm fd}
=&
\int_{-\infty}^{+\infty}
\frac{d\nu}{2\pi}
\left[
-i\nu I
-
K_{{\rm c},j}(\nu)
\right]^{-1}
D_{{\rm c},j}(\nu)
\\
&\times
\left(
\left[
-i\nu I
-
K_{{\rm c},j}(\nu)
\right]^{-1}
\right)^{\dagger}.
\end{split}
\label{eq:app_V_cooling_fd}
\end{equation}
Here $K_{{\rm c},j}(\nu)$ and $D_{{\rm c},j}(\nu)$ are obtained from their drive-phase counterparts by evaluating the phase-dependent optical response at $\phi_j(\nu)$.
The mean field and the radiation-pressure shift are still determined by the drive phase, while the fluctuation response is evaluated at the scattered-field phase.
The corresponding phonon number is extracted from the mechanical block,
\begin{equation}
\bar n_j^{\rm fd}
=
\frac{1}{2}
\left[
(V_{{\rm c},j}^{\rm fd})_{33}
+
(V_{{\rm c},j}^{\rm fd})_{44}
-
1
\right].
\label{eq:app_nbar_cooling_fd}
\end{equation}

\begin{figure}[t]
    \centering
    \includegraphics[width=0.42\textwidth]{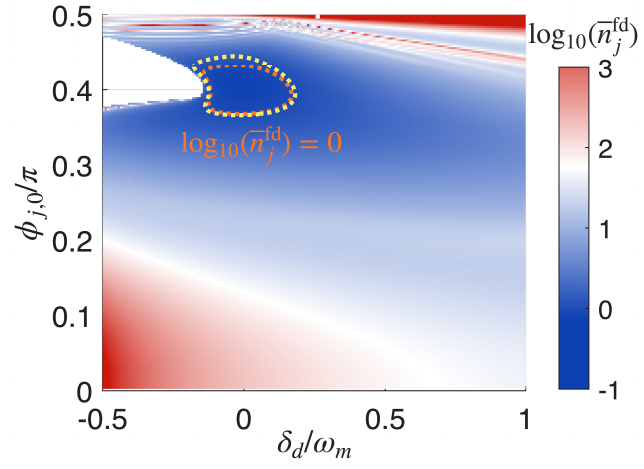}
    \caption{
    Check of the drive-phase approximation for cooling. 
    The color map shows $\log_{10}(\bar n_j^{\rm fd})$, obtained from the frequency-dependent covariance matrix in Eq.~\eqref{eq:app_V_cooling_fd}. 
    The orange dashed contour denotes $\bar n_j^{\rm fd}=1$, while the yellow dashed contour  shows the corresponding boundary obtained under the drive-phase approximation shown in Fig.~\ref{fig:PhaseDep}.
    The cooling window and the ground-state cooling boundary are qualitatively consistent with the drive-phase result shown in Fig.~\ref{fig:PhaseDep}.
    }
    \label{fig:app_cooling_phase_check}
\end{figure}

Figure~\ref{fig:app_cooling_phase_check} shows $\log_{10}(\bar n_j^{\rm fd})$ in the same parameter plane as Fig.~\ref{fig:PhaseDep}. 
The orange dashed contour marks the $\bar n_j^{\rm fd}=1$ boundary obtained from the frequency-dependent scattered-field-phase calculation, while the yellow dashed contour shows the corresponding $\bar n_j=1$ boundary from the drive-phase result in the main text.
The main cooling window and the ground-state cooling boundary remain qualitatively unchanged. 
This confirms that the drive-phase approximation captures the relevant phase-dependent cooling behavior for the parameters considered here.

\section{Minimum mechanical quadrature variance}
\label{app:Vmin}

In this Appendix, we derive the expression for the minimum mechanical quadrature variance used to quantify the mechanical squeezing. 
In the squeezing calculation, the covariance matrix is written in the basis
\begin{equation}
\hat{\mathbf u}_{{\rm s},j}^{T}
=(\delta\hat X_j,
\delta\hat Y_j,
\delta\hat Q,
\delta\hat P).
\label{app_usj}
\end{equation}
The mechanical covariance block is therefore
\begin{equation}
V_m
=
\begin{pmatrix}
(V_{{\rm s},j})_{33} & (V_{{\rm s},j})_{34}\\
(V_{{\rm s},j})_{43} & (V_{{\rm s},j})_{44}
\end{pmatrix}.
\label{app_Vm}
\end{equation}
Since $V_{{\rm s},j}$ is symmetric, one has $(V_{{\rm s},j})_{34}=(V_{{\rm s},j})_{43}$.

The squeezed mechanical quadrature is not necessarily aligned with either $\delta\hat Q$ or $\delta\hat P$. 
Therefore, we consider a general rotated quadrature
\begin{equation}
\delta\hat Q_{\theta}
=
\delta\hat Q\cos\theta
+
\delta\hat P\sin\theta.
\label{app_Qtheta}
\end{equation}
Its variance is
\begin{equation}
\begin{split}
V_{\theta}
&=
\langle \delta\hat Q_{\theta}^{2}\rangle
\\
&=
(V_{{\rm s},j})_{33}\cos^2\theta
+
(V_{{\rm s},j})_{44}\sin^2\theta
\\
&\quad
+
2(V_{{\rm s},j})_{34}
\sin\theta\cos\theta.
\end{split}
\label{app_Vtheta}
\end{equation}
This can be rewritten as
\begin{equation}
\begin{split}
V_{\theta}
=
&\frac{
(V_{{\rm s},j})_{33}
+
(V_{{\rm s},j})_{44}
}{2}
+
\frac{
(V_{{\rm s},j})_{33}
-
(V_{{\rm s},j})_{44}
}{2}
\cos2\theta
\\&+
(V_{{\rm s},j})_{34}\sin2\theta.
\label{app_Vtheta2}
\end{split}
\end{equation}
The minimum over $\theta$ is obtained by choosing the angle such that the last two terms give the largest negative contribution. 
This yields
\begin{equation}
\begin{split}
V_{\min,j}=&\min_{\theta} V_{\theta}
=
\frac{
(V_{{\rm s},j})_{33}
+
(V_{{\rm s},j})_{44}
}{2}
\\&-
\sqrt{
\left[
\frac{
(V_{{\rm s},j})_{33}
-
(V_{{\rm s},j})_{44}
}{2}
\right]^2
+
(V_{{\rm s},j})_{34}^{2}
}.
\label{app_Vmin_derivation}
\end{split}
\end{equation}
The corresponding optimal quadrature angle satisfies
\begin{equation}
\tan 2\theta_{\rm opt}
=
\frac{
2(V_{{\rm s},j})_{34}
}{
(V_{{\rm s},j})_{33}
-
(V_{{\rm s},j})_{44}
},
\label{app_theta_opt}
\end{equation}
with the branch chosen such that $V_\theta$ is minimized.

\section{Check of the center-phase approximation for mechanical squeezing}
\label{app:squeezing_phase_check}

\begin{figure}[t]
    \centering
    \includegraphics[width=0.42\textwidth]{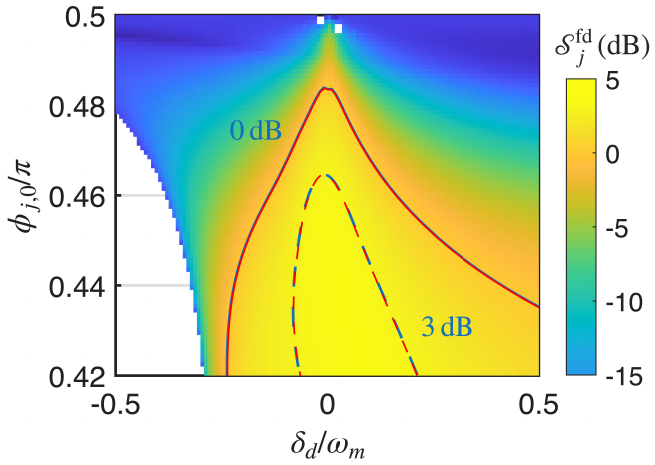}
    \caption{
    Check of the center-phase approximation for mechanical squeezing.
    The color map shows the frequency-dependent squeezing $S_j^{\rm fd}$ obtained from Eq.~\eqref{eq:app_V_squeezing_fd} within the frequency window $|\nu|\leq10\omega_m$.
    The red contours show the corresponding boundaries obtained under the center-phase approximation, as shown in Fig.~\ref{fig:squeezing}, while the corresponding blue contours are obtained from the frequency-dependent calculation.
    The close overlap of the two sets of contours confirms the validity of the center-phase approximation in the strong-squeezing region.
    }
    \label{app_squeezing_fd}
\end{figure}

For the squeezing calculation presented in the main text, the optical fluctuation response is evaluated at the two-tone center phase, i.e., the center-phase approximation
$\phi_j(\nu)\simeq \phi_{j,s}$ is used.
Similarly, we check the validity of this approximation by restoring the frequency-dependent scattered-field phase as in Appendix~\ref{app:cooling_phase_check}, $\phi_j(\nu)=\phi_{j,s}+
\nu L/v_g$,
where $\nu$ is the Fourier frequency measured from the two-tone center frequency.
Accordingly, the parameters entering the fluctuation dynamics become $\nu$-dependent.
The coherent red- and blue-sideband mean fields, and hence $G_{j,-}$ and $G_{j,+}$, are still determined from their respective phases $\phi_{j,-}$ and $\phi_{j,+}$.

To examine the frequency dependence of the optical reservoir around the two-tone center, we evaluate the frequency-dependent correction within a finite fluctuation-frequency window $|\nu|\leq\nu_c$. 
The covariance matrix is written as
\begin{equation}
\begin{split}
V_{{\rm s},j}^{\rm fd}
=
&V_{{\rm s},j}^{\rm cp}
+
\int_{-\nu_c}^{+\nu_c}
\frac{d\nu}{2\pi}
\Bigg\{
\left[
-i\nu I-K_{{\rm s},j}(\nu)
\right]^{-1}
D_{{\rm s},j}(\nu)
\\
&\times
\left(
\left[
-i\nu I-K_{{\rm s},j}(\nu)
\right]^{-1}
\right)^{\dagger}
-\left[
-i\nu I-K_{{\rm s},j}^{\rm cp}
\right]^{-1}
D_{{\rm s},j}^{\rm cp}
\\
&\times
\left(
\left[
-i\nu I-K_{{\rm s},j}^{\rm cp}
\right]^{-1}
\right)^{\dagger}
\Bigg\},
\end{split}
\label{eq:app_V_squeezing_fd}
\end{equation}
where $V_{{\rm s},j}^{\rm cp}$ is the covariance matrix obtained from the center-phase Lyapunov equation in the main text. 
The two terms in the integral are evaluated over the same frequency window, so that their difference isolates the change caused by replacing the center phase $\phi_{j,s}$ with the frequency-dependent phase $\phi_j(\nu)$. 
Here $K_{{\rm s},j}(\nu)$ and $D_{{\rm s},j}(\nu)$ are the corresponding frequency-dependent drift and diffusion matrices, while $K_{{\rm s},j}^{\rm cp}$ and $D_{{\rm s},j}^{\rm cp}$ denote their center-phase counterparts. 
The squeezing level is then obtained from the mechanical block of $V_{{\rm s},j}^{\rm fd}$ as
\begin{equation}
S_j^{\rm fd}
=
-10\log_{10}
\left(
2V_{{\rm min},j}^{\rm fd}
\right),
\label{eq:app_S_squeezing_fd}
\end{equation}
where $V_{{\rm min},j}^{\rm fd}$ follows from Eq.~\eqref{app_Vmin_derivation} with $V_{{\rm s},j}$ replaced by $V_{{\rm s},j}^{\rm fd}$.

Figure~\ref{app_squeezing_fd} shows $S_j^{\rm fd}$ in the strong-squeezing region.
The red solid and dashed contours denote the $S_j=0$ dB and $S_j=3$ dB boundaries obtained under the center-phase approximation, while the corresponding blue contours are obtained from the frequency-dependent calculation.
The two sets of contours almost completely overlap.
This confirms that the center-phase approximation captures the relevant phase-dependent squeezing behavior for the parameters considered here.

The good agreement found in both cases can be understood from the fact that the relevant dynamics are dominated by the optical response around the driving frequency for cooling and the two-tone center frequency for squeezing, while the frequency dependence of the propagation phase over the spectral range that contributes appreciably to the fluctuations produces only a small correction to the final observables.
This allows the  optical response to be represented accurately by the value at the corresponding central frequency for the parameters considered here.

\section{Check of the rotating-wave approximation for mechanical squeezing}
\label{app:rwa_check}

In the main text, mechanical squeezing is calculated within an effective RWA, which leads to the time-independent drift matrix $K_{{\rm s},j}$.
Here we examine the influence of the counter-rotating terms in
the fluctuation dynamics by solving the corresponding
time-periodic covariance equation. The coherent intracavity
amplitudes and the cycle-averaged mechanical displacement,
\begin{equation}
\bar q_{{\rm s},j}
=
\frac{g}{\omega_m}
\left(
|\alpha_{j,-}|^2+|\alpha_{j,+}|^2
\right),
\end{equation}
are kept fixed at their values used in the main-text RWA
treatment. Thus, only the counter-rotating terms in the
fluctuation dynamics are restored.
In this treatment, the drift matrix is written as
\begin{equation}
K_j(t)=K_{{\rm s},j}+K_j^{\rm cr}(t),
\quad
K_j^{\rm cr}(t+T_m)=K_j^{\rm cr}(t),
\label{eq:app_K_periodic}
\end{equation}
where $T_m=2\pi/\omega_m$ and 
the counter-rotating contribution is
\begin{widetext}
\begin{equation}
K_j^{\rm cr}(t)
=
\begin{pmatrix}
0 & 0
& -G_{j,D}\sin(2\omega_m t)/2
& G_{j,D}\cos(2\omega_m t)/2
\\
0 & 0
& G_{j,\Sigma}\cos(2\omega_m t)/2
& G_{j,\Sigma}\sin(2\omega_m t)/2
\\
-G_{j,\Sigma}\sin(2\omega_m t)/2
& G_{j,D}\cos(2\omega_m t)/2
& 0 & 0
\\
G_{j,\Sigma}\cos(2\omega_m t)/2
& G_{j,D}\sin(2\omega_m t)/2
& 0 & 0
\end{pmatrix}.
\label{eq:app_Kcr_explicit}
\end{equation}
\end{widetext}
All its matrix elements oscillate at $2\omega_m$, and its
average over $T_m$ vanishes,
\begin{equation}
\frac{1}{T_m}
\int_0^{T_m}
K_j^{\rm cr}(t)\,dt
=
0.
\label{eq:app_Kcr_average}
\end{equation}
Therefore, the period average of $K_j(t)$ recovers the RWA drift
matrix $K_{{\rm s},j}$ used in the main text.

\begin{figure}[t]
\centering
\includegraphics[width=0.42\textwidth]{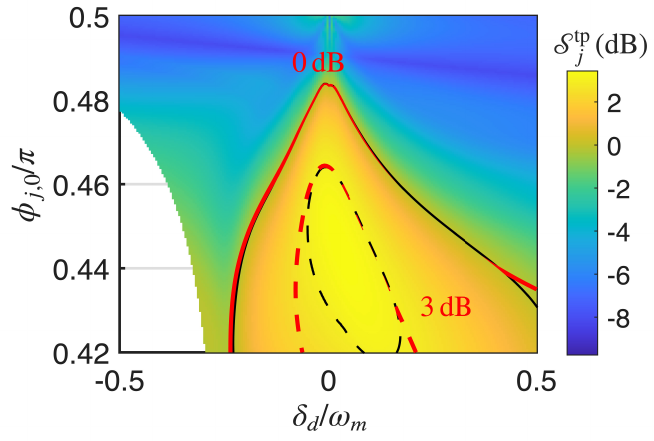}
\caption{
Check of the rotating-wave approximation for squeezing. 
The color map shows $\mathcal S_j^{\rm tp}$ obtained by retaining the counter-rotating terms. 
The black solid and dashed contours denote $\mathcal S_j^{\rm tp}=0\,{\rm dB}$ and $\mathcal S_j^{\rm tp}=3\,{\rm dB}$, respectively. 
The red contours show the corresponding boundaries obtained under the RWA, as shown in Fig.~\ref{fig:squeezing}. 
The optimal squeezed region remains present beyond the RWA, although the RWA overestimates the extent of the squeezed region.}
\label{fig:app_rwa_full_comparison}
\end{figure}

The covariance matrix is obtained from the periodic Lyapunov equation
\begin{equation}
\frac{dV_j(t)}{dt}
=K_j(t)V_j(t)+V_j(t)K_j^T(t)+D_{{\rm s},j},
\label{eq:app_periodic_cov}
\end{equation}
with the periodic steady-state condition
\begin{equation}
V_j^{\rm ss}(t+T_m)=V_j^{\rm ss}(t).
\label{eq:app_periodic_ss}
\end{equation}
The squeezing is extracted from the period-averaged mechanical covariance,
\begin{equation}
\overline{V}_{m,j}^{\rm tp}
=
\frac{1}{T_m}
\int_0^{T_m}
V_{m,j}^{\rm ss}(t)\,dt ,
\label{eq:app_Vm_full_avg}
\end{equation}
as
\begin{equation}
\mathcal S_j^{\rm tp}
=
-10\log_{10}
\left[
2\lambda_{\min}
\left(
\overline{V}_{m,j}^{\rm tp}
\right)
\right].
\label{eq:app_S_full}
\end{equation}

Figure~\ref{fig:app_rwa_full_comparison}
shows the time-periodic squeezing 
$\mathcal S_j^{\rm tp}$ obtained by retaining the counter-rotating terms.
The black solid and dashed contours indicate $\mathcal S_j^{\rm tp}=0\,{\rm dB}$ and $\mathcal S_j^{\rm tp}=3\,{\rm dB}$, respectively, while the red contours indicate the corresponding RWA thresholds shown in Fig.~\ref{fig:squeezing}.
The comparison shows that the RWA captures the location of the optimal squeezed region, although it overestimates the extent of the squeezed region. 
Thus, the RWA provides a useful physical picture, while the time-periodic calculation confirms that the main squeezed region remains present when the counter-rotating terms are included.

\bibliography{references}

@article{small1,
  title = {{Tunable delay line with interacting whispering-gallery-mode resonators}},
  volume = {29},
  ISSN = {1539-4794},
  url = {http://dx.doi.org/10.1364/OL.29.000626},
  DOI = {10.1364/ol.29.000626},
  number = {6},
  journal = {Opt. Lett.},
  publisher = {Optica Publishing Group},
  author = {Maleki,  L. and Matsko,  A. B. and Savchenkov,  A. A. and Ilchenko,  V. S.},
  year = {2004},
  month = mar,
  pages = {626}
}

@article{small2,
  title = {{Coupling Whispering-Gallery-Mode Microcavities With Modal Coupling Mechanism}},
  volume = {44},
  ISSN = {0018-9197},
  url = {http://dx.doi.org/10.1109/JQE.2008.2002088},
  DOI = {10.1109/jqe.2008.2002088},
  number = {11},
  journal = {IEEE J. Quantum Electron.},
  publisher = {Institute of Electrical and Electronics Engineers (IEEE)},
  author = {Xiao,  Yun-Feng and Min,  Bumki and Jiang,  Xiaoshun and Dong,  Chun-Hua and Yang,  Lan},
  year = {2008},
  month = nov,
  pages = {1065–1070}
}

@article{Gardiner1985,
  title = {{Input and output in damped quantum systems: Quantum stochastic differential equations and the master equation}},
  volume = {31},
  ISSN = {0556-2791},
  url = {http://dx.doi.org/10.1103/PhysRevA.31.3761},
  DOI = {10.1103/physreva.31.3761},
  number = {6},
  journal = {Phys. Rev. A},
  publisher = {American Physical Society (APS)},
  author = {Gardiner,  C. W. and Collett,  M. J.},
  year = {1985},
  month = Jun,
  pages = {3761–3774}
}

@article{anton2014,
  title = {{Designing frequency-dependent relaxation rates and Lamb shifts for a giant artificial atom}},
  volume = {90},
  ISSN = {1094-1622},
  url = {http://dx.doi.org/10.1103/PhysRevA.90.013837},
  DOI = {10.1103/physreva.90.013837},
  number = {1},
  journal = {Phys. Rev. A},
  publisher = {American Physical Society (APS)},
  author = {Frisk Kockum,  Anton and Delsing,  Per and Johansson,  G\"{o}ran},
  year = {2014},
  month = July,
  pages = {013837}
}

@article{WilsonRae2007,
  title = {{Theory of Ground State Cooling of a Mechanical Oscillator Using Dynamical Backaction}},
  volume = {99},
  ISSN = {1079-7114},
  url = {http://dx.doi.org/10.1103/PhysRevLett.99.093901},
  DOI = {10.1103/physrevlett.99.093901},
  number = {9},
  journal = {Phys. Rev. Lett.},
  publisher = {American Physical Society (APS)},
  author = {Wilson-Rae,  I. and Nooshi,  N. and Zwerger,  W. and Kippenberg,  T. J.},
  year = {2007},
  month = Aug,
  pages = {093901}
}

@article{Marquardt2007,
  title = {{Quantum Theory of Cavity-Assisted Sideband Cooling of Mechanical Motion}},
  volume = {99},
  ISSN = {1079-7114},
  url = {http://dx.doi.org/10.1103/PhysRevLett.99.093902},
  DOI = {10.1103/physrevlett.99.093902},
  number = {9},
  journal = {Phys. Rev. Lett.},
  publisher = {American Physical Society (APS)},
  author = {Marquardt,  Florian and Chen,  Joe P. and Clerk,  A. A. and Girvin,  S. M.},
  year = {2007},
  month = Aug,
  pages = {093902}
}

@article{Aspelmeyer2014,
  title = {{Cavity optomechanics}},
  volume = {86},
  ISSN = {1539-0756},
  url = {http://dx.doi.org/10.1103/RevModPhys.86.1391},
  DOI = {10.1103/revmodphys.86.1391},
  number = {4},
  journal = {Rev. Mod. Phys.},
  publisher = {American Physical Society (APS)},
  author = {Aspelmeyer,  Markus and Kippenberg,  Tobias J. and Marquardt,  Florian},
  year = {2014},
  month = Dec,
  pages = {1391–1452}
}

@article{Genes2008,
  title = {{Ground-state cooling of a micromechanical oscillator: Comparing cold damping and cavity-assisted cooling schemes}},
  volume = {77},
  ISSN = {1094-1622},
  url = {http://dx.doi.org/10.1103/PhysRevA.77.033804},
  DOI = {10.1103/physreva.77.033804},
  number = {3},
  journal = {Phys. Rev. A},
  publisher = {American Physical Society (APS)},
  author = {Genes,  C. and Vitali,  D. and Tombesi,  P. and Gigan,  S. and Aspelmeyer,  M.},
  year = {2008},
  month = Mar,
  pages = {033804}
}

@article{Kronwald2013,
  title = {{Arbitrarily large steady-state bosonic squeezing via dissipation}},
  volume = {88},
  ISSN = {1094-1622},
  url = {http://dx.doi.org/10.1103/PhysRevA.88.063833},
  DOI = {10.1103/physreva.88.063833},
  number = {6},
  journal = {Phys. Rev. A},
  publisher = {American Physical Society (APS)},
  author = {Kronwald,  Andreas and Marquardt,  Florian and Clerk,  Aashish A.},
  year = {2013},
  month = Dec,
  pages = {063833}
}

@article{Maayani2018Nature,
  title = {{Flying couplers above spinning resonators generate irreversible refraction}},
  volume = {558},
  ISSN = {1476-4687},
  url = {http://dx.doi.org/10.1038/s41586-018-0245-5},
  DOI = {10.1038/s41586-018-0245-5},
  number = {7711},
  journal = {Nature},
  publisher = {Springer Science and Business Media LLC},
  author = {Maayani,  Shai and Dahan,  Raphael and Kligerman,  Yuri and Moses,  Eduard and Hassan,  Absar U. and Jing,  Hui and Nori,  Franco and Christodoulides,  Demetrios N. and Carmon,  Tal},
  year = {2018},
  month = June,
  pages = {569–572}
}

@article{Lu2017PRJ,
   title = {{Optomechanically induced transparency in a spinning resonator}},
  volume = {5},
  ISSN = {2327-9125},
  url = {http://dx.doi.org/10.1364/PRJ.5.000367},
  DOI = {10.1364/prj.5.000367},
  number = {4},
  journal = {Photonics Res.},
  publisher = {Optica Publishing Group},
  author = {L\"{u},  Hao and Jiang,  Yajing and Wang,  Yu-Zhu and Jing,  Hui},
  year = {2017},
  month = July,
  pages = {367}
}

@article{Jiang2018PRAppl,
  title = {{Nonreciprocal Phonon Laser}},
  volume = {10},
  ISSN = {2331-7019},
  url = {http://dx.doi.org/10.1103/PhysRevApplied.10.064037},
  DOI = {10.1103/physrevapplied.10.064037},
  number = {6},
  journal = {Phys. Rev. Appl.},
  publisher = {American Physical Society (APS)},
  author = {Jiang,  Y. and Maayani,  S. and Carmon,  T. and Nori,  Franco and Jing,  H.},
  year = {2018},
  month = Dec,
  pages = {064037}
}

@article{Huang2018PRL,
  title = {{Nonreciprocal Photon Blockade}},
  volume = {121},
  ISSN = {1079-7114},
  url = {http://dx.doi.org/10.1103/PhysRevLett.121.153601},
  DOI = {10.1103/physrevlett.121.153601},
  number = {15},
  journal = {Phys. Rev. Lett.},
  publisher = {American Physical Society (APS)},
  author = {Huang,  Ran and Miranowicz,  Adam and Liao,  Jie-Qiao and Nori,  Franco and Jing,  Hui},
  year = {2018},
  month = Oct,
  pages = {153601}
}

@article{Jiao2020PRL,
  title = {{Nonreciprocal Optomechanical Entanglement against Backscattering Losses}},
  volume = {125},
  ISSN = {1079-7114},
  url = {http://dx.doi.org/10.1103/PhysRevLett.125.143605},
  DOI = {10.1103/physrevlett.125.143605},
  number = {14},
  journal = {Phys. Rev. Lett.},
  publisher = {American Physical Society (APS)},
  author = {Jiao,  Ya-Feng and Zhang,  Sheng-Dian and Zhang,  Yan-Lei and Miranowicz,  Adam and Kuang,  Le-Man and Jing,  Hui},
  year = {2020},
  month = Oct,
  pages = {143605}
}

@article{Jalas2013,
  title = {{What is — and what is not — an optical isolator}},
  volume = {7},
  ISSN = {1749-4893},
  url = {http://dx.doi.org/10.1038/nphoton.2013.185},
  DOI = {10.1038/nphoton.2013.185},
  number = {8},
  journal = {Nat. Photonics},
  publisher = {Springer Science and Business Media LLC},
  author = {Jalas,  Dirk and Petrov,  Alexander and Eich,  Manfred and Freude,  Wolfgang and Fan,  Shanhui and Yu,  Zongfu and Baets,  Roel and Popović,  Miloš and Melloni,  Andrea and Joannopoulos,  John D. and Vanwolleghem,  Mathias and Doerr,  Christopher R. and Renner,  Hagen},
  year = {2013},
  month = July,
  pages = {579–582}
}

@article{Lodahl2017,
  title = {{Chiral quantum optics}},
  volume = {541},
  ISSN = {1476-4687},
  url = {http://dx.doi.org/10.1038/nature21037},
  DOI = {10.1038/nature21037},
  number = {7638},
  journal = {Nature},
  publisher = {Springer Science and Business Media LLC},
  author = {Lodahl,  Peter and Mahmoodian,  Sahand and Stobbe,  Søren and Rauschenbeutel,  Arno and Schneeweiss,  Philipp and Volz,  J\"{u}rgen and Pichler,  Hannes and Zoller,  Peter},
  year = {2017},
  month = Jan,
  pages = {473–480}
}

@article{Caloz2018,
  title = {{Electromagnetic Nonreciprocity}},
  volume = {10},
  ISSN = {2331-7019},
  url = {http://dx.doi.org/10.1103/PhysRevApplied.10.047001},
  DOI = {10.1103/physrevapplied.10.047001},
  number = {4},
  journal = {Phys. Rev. Appl.},
  publisher = {American Physical Society (APS)},
  author = {Caloz,  Christophe and Alù,  Andrea and Tretyakov,  Sergei and Sounas,  Dimitrios and Achouri,  Karim and Deck-Léger,  Zoé-Lise},
  year = {2018},
  month = Oct,
  pages = {047001}
}

@article{Kannan2020,
  title = {{Waveguide quantum electrodynamics with superconducting artificial giant atoms}},
  volume = {583},
  ISSN = {1476-4687},
  url = {http://dx.doi.org/10.1038/s41586-020-2529-9},
  DOI = {10.1038/s41586-020-2529-9},
  number = {7818},
  journal = {Nature},
  publisher = {Springer Science and Business Media LLC},
  author = {Kannan,  Bharath and Ruckriegel,  Max J. and Campbell,  Daniel L. and Frisk Kockum,  Anton and Braum\"{u}ller,  Jochen and Kim,  David K. and Kjaergaard,  Morten and Krantz,  Philip and Melville,  Alexander and Niedzielski,  Bethany M. and Veps\"{a}l\"{a}inen,  Antti and Winik,  Roni and Yoder,  Jonilyn L. and Nori,  Franco and Orlando,  Terry P. and Gustavsson,  Simon and Oliver,  William D.},
  year = {2020},
  month = July,
  pages = {775–779}
}

@article{Zhang2021PRA,
  title = {{Nonreciprocal chaos in a spinning optomechanical resonator}},
  volume = {104},
  ISSN = {2469-9934},
  url = {http://dx.doi.org/10.1103/PhysRevA.104.033522},
  DOI = {10.1103/physreva.104.033522},
  number = {3},
  journal = {Phys. Rev. A},
  publisher = {American Physical Society (APS)},
  author = {Zhang,  Deng-Wei and Zheng,  Li-Li and You,  Cai and Hu,  Chang-Sheng and Wu,  Ying and L\"{u},  Xin-You},
  year = {2021},
  month = Sept,
  pages = {033522}
}

@article{Yuan2023OE,
  title = {{Optical noise-resistant nonreciprocal phonon blockade in a spinning optomechanical resonator}},
  volume = {31},
  ISSN = {1094-4087},
  url = {http://dx.doi.org/10.1364/OE.492209},
  DOI = {10.1364/oe.492209},
  number = {12},
  journal = {Opt. Express},
  publisher = {Optica Publishing Group},
  author = {Yuan,  Ning and He,  Shuang and Li,  Shi-Yan and Wang,  Nan and Zhu,  Ai-Dong},
  year = {2023},
  month = May,
  pages = {20160}
}

@article{Guo2023PRA,
  title = {{Nonreciprocal mechanical squeezing in a spinning cavity optomechanical system via pump modulation}},
  volume = {108},
  ISSN = {2469-9934},
  url = {http://dx.doi.org/10.1103/PhysRevA.108.033515},
  DOI = {10.1103/physreva.108.033515},
  number = {3},
  journal = {Phys. Rev. A},
  publisher = {American Physical Society (APS)},
  author = {Guo,  Qi and Zhou,  Ke-Xin and Bai,  Cheng-Hua and Zhang,  Yuchi and Li,  Gang and Zhang,  Tiancai},
  year = {2023},
  month = Sept,
  pages = {033515}
}

@inbook{FriskKockum2020,
  title = {{Quantum Optics with Giant Atoms—the First Five Years}},
  ISBN = {9789811551918},
  ISSN = {2198-3518},
  url = {http://dx.doi.org/10.1007/978-981-15-5191-8_12},
  DOI = {10.1007/978-981-15-5191-8_12},
  booktitle = {International Symposium on Mathematics, Quantum Theory, and Cryptography},
  publisher = {Springer Singapore},
  author = {Frisk Kockum,  Anton},
  year = {2020},
  month = Oct,
  pages = {125–146}
}

@article{Gustafsson2014,
  title = {{Propagating phonons coupled to an artificial atom}},
  volume = {346},
  ISSN = {1095-9203},
  url = {http://dx.doi.org/10.1126/science.1257219},
  DOI = {10.1126/science.1257219},
  number = {6206},
  journal = {Science},
  publisher = {American Association for the Advancement of Science (AAAS)},
  author = {Gustafsson,  Martin V. and Aref,  Thomas and Kockum,  Anton Frisk and Ekstr\"{o}m,  Maria K. and Johansson,  G\"{o}ran and Delsing,  Per},
  year = {2014},
  month = Oct,
  pages = {207–211}
}

@article{Zhu2022FrontPhys,
  title = {{Giant-Cavity-Based Quantum Sensors With Enhanced Performance}},
  volume = {10},
  ISSN = {2296-424X},
  url = {http://dx.doi.org/10.3389/fphy.2022.896596},
  DOI = {10.3389/fphy.2022.896596},
  journal = {Front. Phys.},
  publisher = {Frontiers Media SA},
  author = {Zhu,  Y. T. and Wu,  R. B. and Peng,  Z. H. and Xue,  Shibei},
  year = {2022},
  month = June,
  pages = {896596}
}

@article{Du2021OE,
  title = {{Controllable optical response and tunable sensing based on self interference in waveguide QED systems}},
  volume = {29},
  ISSN = {1094-4087},
  url = {http://dx.doi.org/10.1364/OE.412996},
  DOI = {10.1364/oe.412996},
  number = {3},
  journal = {Opt. Express},
  publisher = {Optica Publishing Group},
  author = {Du,  Lei and Wang,  Zhihai and Li,  Yong},
  year = {2021},
  month = Jan,
  pages = {3038}
}

@article{Liu2022,
  title = {{Nonreciprocal Waveguide-QED for Spinning Cavities with Multiple Coupling Points}},
  volume = {10},
  ISSN = {2296-424X},
  url = {http://dx.doi.org/10.3389/fphy.2022.894115},
  DOI = {10.3389/fphy.2022.894115},
  journal = {Front. Phys.},
  publisher = {Frontiers Media SA},
  author = {Liu,  Wenxiao and Lin,  Yafen and Li,  Jiaqi and Wang,  Xin},
  year = {2022},
  month = Apr,
  pages = {894115}
}

@article{Wang2022,
  title = {{Giant spin ensembles in waveguide magnonics}},
  volume = {13},
  ISSN = {2041-1723},
  url = {http://dx.doi.org/10.1038/s41467-022-35174-9},
  DOI = {10.1038/s41467-022-35174-9},
  number = {1},
  journal = {Nat. Commun.},
  publisher = {Springer Science and Business Media LLC},
  author = {Wang,  Zi-Qi and Wang,  Yi-Pu and Yao,  Jiguang and Shen,  Rui-Chang and Wu,  Wei-Jiang and Qian,  Jie and Li,  Jie and Zhu,  Shi-Yao and You,  J. Q.},
  year = {2022},
  month = Dec,
  pages = {7580}
}

@article{Chang2025,
  title = {{Non-Markovian multiphoton chiral dynamics with giant systems}},
  volume = {8},
  ISSN = {2399-3650},
  url = {http://dx.doi.org/10.1038/s42005-025-02293-w},
  DOI = {10.1038/s42005-025-02293-w},
  number = {1},
  journal = {Commun. Phys.},
  publisher = {Springer Science and Business Media LLC},
  author = {Chang,  Yue},
  year = {2025},
  month = Sept,
  pages = {385}
}

@article{Ruesink2016,
  title = {{Nonreciprocity and magnetic-free isolation based on optomechanical interactions}},
  volume = {7},
  ISSN = {2041-1723},
  url = {http://dx.doi.org/10.1038/ncomms13662},
  DOI = {10.1038/ncomms13662},
  number = {1},
  journal = {Nat. Commun.},
  publisher = {Springer Science and Business Media LLC},
  author = {Ruesink,  Freek and Miri,  Mohammad-Ali and Alù,  Andrea and Verhagen,  Ewold},
  year = {2016},
  month = Nov,
  pages = {13662}
}

@article{Andersson2019,
  title = {{Non-exponential decay of a giant artificial atom}},
  volume = {15},
  ISSN = {1745-2481},
  url = {http://dx.doi.org/10.1038/s41567-019-0605-6},
  DOI = {10.1038/s41567-019-0605-6},
  number = {11},
  journal = {Nat. Phys.},
  publisher = {Springer Science and Business Media LLC},
  author = {Andersson,  Gustav and Suri,  Baladitya and Guo,  Lingzhen and Aref,  Thomas and Delsing,  Per},
  year = {2019},
  month = Aug,
  pages = {1123–1127}
}

@article{Wan2018,
  title = {{Experimental demonstration of dissipative sensing in a self-interference microring resonator}},
  volume = {6},
  ISSN = {2327-9125},
  url = {http://dx.doi.org/10.1364/PRJ.6.000681},
  DOI = {10.1364/prj.6.000681},
  number = {7},
  journal = {Photonics Res.},
  publisher = {Optica Publishing Group},
  author = {Wan,  Shuai and Niu,  Rui and Ren,  Hong-Liang and Zou,  Chang-Ling and Guo,  Guang-Can and Dong,  Chun-Hua},
  year = {2018},
  month = June,
  pages = {681}
}

@article{Mazzei2007,
  title = {{Controlled Coupling of Counterpropagating Whispering-Gallery Modes by a Single Rayleigh Scatterer: A Classical Problem in a Quantum Optical Light}},
  volume = {99},
  ISSN = {1079-7114},
  url = {http://dx.doi.org/10.1103/PhysRevLett.99.173603},
  DOI = {10.1103/physrevlett.99.173603},
  number = {17},
  journal = {Phys. Rev. Lett.},
  publisher = {American Physical Society (APS)},
  author = {Mazzei,  A. and G\"{o}tzinger,  S. and de S. Menezes,  L. and Zumofen,  G. and Benson,  O. and Sandoghdar,  V.},
  year = {2007},
  month = Oct,
  pages = {173603}
}

@article{Schliesser2008,
  title = {{Resolved-sideband cooling of a micromechanical oscillator}},
  volume = {4},
  ISSN = {1745-2481},
  url = {http://dx.doi.org/10.1038/nphys939},
  DOI = {10.1038/nphys939},
  number = {5},
  journal = {Nat. Phys.},
  publisher = {Springer Science and Business Media LLC},
  author = {Schliesser,  A. and Rivière,  R. and Anetsberger,  G. and Arcizet,  O. and Kippenberg,  T. J.},
  year = {2008},
  month = Apr,
  pages = {415–419}
}

@article{Teufel2011,
  title = {{Sideband cooling of micromechanical motion to the quantum ground state}},
  volume = {475},
  ISSN = {1476-4687},
  url = {http://dx.doi.org/10.1038/nature10261},
  DOI = {10.1038/nature10261},
  number = {7356},
  journal = {Nature},
  publisher = {Springer Science and Business Media LLC},
  author = {Teufel,  J. D. and Donner,  T. and Li,  Dale and Harlow,  J. W. and Allman,  M. S. and Cicak,  K. and Sirois,  A. J. and Whittaker,  J. D. and Lehnert,  K. W. and Simmonds,  R. W.},
  year = {2011},
  month = July,
  pages = {359–363}
}

@article{Chan2011,
  title = {{Laser cooling of a nanomechanical oscillator into its quantum ground state}},
  volume = {478},
  ISSN = {1476-4687},
  url = {http://dx.doi.org/10.1038/nature10461},
  DOI = {10.1038/nature10461},
  number = {7367},
  journal = {Nature},
  publisher = {Springer Science and Business Media LLC},
  author = {Chan,  Jasper and Alegre,  T. P. Mayer and Safavi-Naeini,  Amir H. and Hill,  Jeff T. and Krause,  Alex and Gr\"{o}blacher,  Simon and Aspelmeyer,  Markus and Painter,  Oskar},
  year = {2011},
  month = Oct,
  pages = {89–92}
}

@article{Mari2009,
  title = {{Gently Modulating Optomechanical Systems}},
  volume = {103},
  ISSN = {1079-7114},
  url = {http://dx.doi.org/10.1103/PhysRevLett.103.213603},
  DOI = {10.1103/physrevlett.103.213603},
  number = {21},
  journal = {Phys. Rev. Lett.},
  publisher = {American Physical Society (APS)},
  author = {Mari,  A. and Eisert,  J.},
  year = {2009},
  month = Nov,
  pages = {213603}
}

@article{Wollman2015,
  title = {{Quantum squeezing of motion in a mechanical resonator}},
  volume = {349},
  ISSN = {1095-9203},
  url = {http://dx.doi.org/10.1126/science.aac5138},
  DOI = {10.1126/science.aac5138},
  number = {6251},
  journal = {Science},
  publisher = {American Association for the Advancement of Science (AAAS)},
  author = {Wollman,  E. E. and Lei,  C. U. and Weinstein,  A. J. and Suh,  J. and Kronwald,  A. and Marquardt,  F. and Clerk,  A. A. and Schwab,  K. C.},
  year = {2015},
  month = Aug,
  pages = {952–955}
}

@article{Lei2016,
  title = {{Quantum Nondemolition Measurement of a Quantum Squeezed State Beyond the 3 dB Limit}},
  volume = {117},
  ISSN = {1079-7114},
  url = {http://dx.doi.org/10.1103/PhysRevLett.117.100801},
  DOI = {10.1103/physrevlett.117.100801},
  number = {10},
  journal = {Phys. Rev. Lett.},
  publisher = {American Physical Society (APS)},
  author = {Lei,  C. U. and Weinstein,  A. J. and Suh,  J. and Wollman,  E. E. and Kronwald,  A. and Marquardt,  F. and Clerk,  A. A. and Schwab,  K. C.},
  year = {2016},
  month = Aug,
  pages = {100801}
}

@article{Manipatruni2009PRL,
  title = {{Optical Nonreciprocity in Optomechanical Structures}},
  volume = {102},
  ISSN = {1079-7114},
  url = {http://dx.doi.org/10.1103/PhysRevLett.102.213903},
  DOI = {10.1103/physrevlett.102.213903},
  number = {21},
  journal = {Phys. Rev. Lett.},
  publisher = {American Physical Society (APS)},
  author = {Manipatruni,  Sasikanth and Robinson,  Jacob T. and Lipson,  Michal},
  year = {2009},
  month = May,
  pages = {213903}
}

@article{Fang2017NatPhys,
  title = {{Generalized non-reciprocity in an optomechanical circuit via synthetic magnetism and reservoir engineering}},
  volume = {13},
  ISSN = {1745-2481},
  url = {http://dx.doi.org/10.1038/nphys4009},
  DOI = {10.1038/nphys4009},
  number = {5},
  journal = {Nat. Phys.},
  publisher = {Springer Science and Business Media LLC},
  author = {Fang,  Kejie and Luo,  Jie and Metelmann,  Anja and Matheny,  Matthew H. and Marquardt,  Florian and Clerk,  Aashish A. and Painter,  Oskar},
  year = {2017},
  month = Jan,
  pages = {465–471}
}

@article{Guo2017,
  title = {{Giant acoustic atom: A single quantum system with a deterministic time delay}},
  volume = {95},
  ISSN = {2469-9934},
  url = {http://dx.doi.org/10.1103/PhysRevA.95.053821},
  DOI = {10.1103/physreva.95.053821},
  number = {5},
  journal = {Phys. Rev. A},
  publisher = {American Physical Society (APS)},
  author = {Guo,  Lingzhen and Grimsmo,  Arne and Kockum,  Anton Frisk and Pletyukhov,  Mikhail and Johansson,  G\"{o}ran},
  year = {2017},
  month = May,
  pages = {053821}
}

@article{Du2022,
  title = {{Giant atoms with time-dependent couplings}},
  volume = {4},
  ISSN = {2643-1564},
  url = {http://dx.doi.org/10.1103/PhysRevResearch.4.023198},
  DOI = {10.1103/physrevresearch.4.023198},
  number = {2},
  journal = {Phys. Rev. Res.},
  publisher = {American Physical Society (APS)},
  author = {Du,  Lei and Chen,  Yao-Tong and Zhang,  Yan and Li,  Yong},
  year = {2022},
  month = June,
  pages = {023198}
}

@article{Wang2026,
  title = {{Unidirectional exceptional point of reflectionless states in a magnonic mirror array}},
  volume = {12},
  ISSN = {2375-2548},
  url = {http://dx.doi.org/10.1126/sciadv.aea6000},
  DOI = {10.1126/sciadv.aea6000},
  number = {10},
  journal = {Sci. Adv.},
  publisher = {American Association for the Advancement of Science (AAAS)},
  author = {Wang,  Zi-Qi and Peng,  Yuan-Peng and Wang,  Yi-Pu and You,  J. Q.},
  year = {2026},
  month = Mar,
  pages = {eaea6000}
}

@article{Wang2021,
  title = {{Tunable Chiral Bound States with Giant Atoms}},
  volume = {126},
  ISSN = {1079-7114},
  url = {http://dx.doi.org/10.1103/PhysRevLett.126.043602},
  DOI = {10.1103/physrevlett.126.043602},
  number = {4},
  journal = {Phys. Rev. Lett.},
  publisher = {American Physical Society (APS)},
  author = {Wang,  Xin and Liu,  Tao and Kockum,  Anton Frisk and Li,  Hong-Rong and Nori,  Franco},
  year = {2021},
  month = Jan,
  pages = {043602}
}

@article{Zhao2020,
  title = {{Single-photon scattering and bound states in an atom-waveguide system with two or multiple coupling points}},
  volume = {101},
  ISSN = {2469-9934},
  url = {http://dx.doi.org/10.1103/PhysRevA.101.053855},
  DOI = {10.1103/physreva.101.053855},
  number = {5},
  journal = {Phys. Rev. A},
  publisher = {American Physical Society (APS)},
  author = {Zhao,  Wei and Wang,  Zhihai},
  year = {2020},
  month = May,
  pages = {053855}
}

@article{Guo2020,
  title = {{Oscillating bound states for a giant atom}},
  volume = {2},
  ISSN = {2643-1564},
  url = {http://dx.doi.org/10.1103/PhysRevResearch.2.043014},
  DOI = {10.1103/physrevresearch.2.043014},
  number = {4},
  journal = {Phys. Rev. Res.},
  publisher = {American Physical Society (APS)},
  author = {Guo,  Lingzhen and Kockum,  Anton Frisk and Marquardt,  Florian and Johansson,  G\"{o}ran},
  year = {2020},
  month = Oct,
  pages = {043014}
}

@article{Du2021,
  title = {{Single-photon frequency conversion via a giant $\Lambda$-type atom}},
  volume = {104},
  ISSN = {2469-9934},
  url = {http://dx.doi.org/10.1103/PhysRevA.104.023712},
  DOI = {10.1103/physreva.104.023712},
  number = {2},
  journal = {Phys. Rev. A},
  publisher = {American Physical Society (APS)},
  author = {Du,  Lei and Li,  Yong},
  year = {2021},
  month = Aug,
  pages = {023712}
}

@article{Wang2019,
  title = {{Phase-controlled single-photon nonreciprocal transmission in a one-dimensional waveguide}},
  volume = {100},
  ISSN = {2469-9934},
  url = {http://dx.doi.org/10.1103/PhysRevA.100.053809},
  DOI = {10.1103/physreva.100.053809},
  number = {5},
  journal = {Phys. Rev. A},
  publisher = {American Physical Society (APS)},
  author = {Wang,  Zhihai and Du,  Lei and Li,  Yong and Liu,  Yu-xi},
  year = {2019},
  month = Nov,
  pages = {053809}
}

@article{Chen2022,
  title = {{Nonreciprocal and chiral single-photon scattering for giant atoms}},
  volume = {5},
  ISSN = {2399-3650},
  url = {http://dx.doi.org/10.1038/s42005-022-00991-3},
  DOI = {10.1038/s42005-022-00991-3},
  number = {1},
  journal = {Commun. Phys.},
  publisher = {Springer Science and Business Media LLC},
  author = {Chen,  Yao-Tong and Du,  Lei and Guo,  Lingzhen and Wang,  Zhihai and Zhang,  Yan and Li,  Yong and Wu,  Jin-Hui},
  year = {2022},
  month = Aug,
  pages = {215}
}

@article{Cai2021,
title = {{Coherent single-photon scattering spectra for a giant-atom waveguide-QED system beyond the dipole approximation}},
volume = {104},
ISSN = {2469-9934},
url = {http://dx.doi.org/10.1103/PhysRevA.104.033710},
DOI = {10.1103/physreva.104.033710},
number = {3},
journal = {Phys. Rev. A},
publisher = {American Physical Society (APS)},
author = {Cai, Q. Y. and Jia, W. Z.},
year = {2021},
month = sep,
pages = {033710}
}

@article{Vega2021,
title = {{Qubit-photon bound states in topological waveguides with long-range hoppings}},
volume = {104},
ISSN = {2469-9934},
url = {http://dx.doi.org/10.1103/PhysRevA.104.053522},
DOI = {10.1103/physreva.104.053522},
number = {5},
journal = {Phys. Rev. A},
publisher = {American Physical Society (APS)},
author = {Vega, C. and Bello, M. and Porras, D. and Gonz{'a}lez-Tudela, A.},
year = {2021},
month = nov,
pages = {053522}
}

@article{Lim2023,
title = {{Oscillating bound states in non-Markovian photonic lattices}},
volume = {107},
ISSN = {2469-9934},
url = {http://dx.doi.org/10.1103/PhysRevA.107.023716},
DOI = {10.1103/physreva.107.023716},
number = {2},
journal = {Phys. Rev. A},
publisher = {American Physical Society (APS)},
author = {Lim, Kian Hwee and Mok, Wai-Keong and Kwek, Leong-Chuan},
year = {2023},
month = feb,
pages = {023716}
}

@article{Xiao2022,
title = {{Bound state in a giant atom-modulated resonators system}},
volume = {8},
ISSN = {2056-6387},
url = {http://dx.doi.org/10.1038/s41534-022-00591-7},
DOI = {10.1038/s41534-022-00591-7},
number = {1},
journal = {npj Quantum Inf.},
publisher = {Springer Nature},
author = {Xiao, Han and Wang, Luojia and Li, Zheng-Hong and Chen, Xianfeng and Yuan, Luqi},
year = {2022},
month = jul,
pages = {80}
}

@article{Du2021prr,
title = {{Nonreciprocal frequency conversion with chiral $\Lambda$-type atoms}},
volume = {3},
ISSN = {2643-1564},
url = {http://dx.doi.org/10.1103/PhysRevResearch.3.043226},
DOI = {10.1103/physrevresearch.3.043226},
number = {4},
journal = {Phys. Rev. Res.},
publisher = {American Physical Society (APS)},
author = {Du, Lei and Chen, Yao-Tong and Li, Yong},
year = {2021},
month = dec,
pages = {043226}
}

\end{document}